\documentclass[11pt]{article}
\usepackage{amsmath,amsthm,latexsym,amssymb,amsfonts,epsfig,bbm,textgreek}
\usepackage[T1]{fontenc} 
\usepackage[utf8]{inputenc} 
\usepackage{psfrag,bbm}

\newtheorem{theorem}{Theorem}

\begin{document}
\title{{\sf The kinematical Setup of Quantum Geometry: A Brief Review}}
\author{
{\sf K. Giesel$^1$\thanks{{\sf 
kristina.giesel@gravity.fau.de}}}
\\
{\sf $^1$ Institute for Quantum Gravity (IQG)} \\ {FAU Erlangen -- N\"urnberg,}\\
{\sf Staudtstr. 7, 91058 Erlangen, Germany}\\
}
\date{{\small\sf \today}}
\maketitle
\begin{abstract}
In this article we present a brief introduction to the kinematical setup that underlies the quantization used in loop quantum gravity. This review has been published as a chapter in the monograph {\it "Loop Quantum Gravity: The First 30 Years"}, edited by Abhay Ashtekar and Jorge Pullin, that was recently published in the series {\it "100 Years of General Relativity"} \cite{Book}.
\end{abstract}

\section{Canonical Quantization of General Relativity}
With their seminal work in 1960 on the canonical formulation of general relativity, nowadays called the ADM-formalism, Arnowitt, Deser and Misner \cite{Deser:1960zzc} provided the background for research which focused on the question how general relativity can be quantized using the technique of canonical quantization. In the covariant formulation of general relativity the elementary variable is a Lorentzian metric  $g_{\mu\nu}$, where we use the signature $(3,1)$, on a four-dimensional differentiable manifold ${\cal M}$. The equation of motion for $g_{\mu\nu}$ are given by Einstein's equations and encode the dynamics of general relativity.  In the context of the ADM-formalism the four-dimensional space time $({\cal M}, g_{\mu\nu})$ is replaced by a 3+1-dimensional picture, using that for globally hyperbolic space times their topology is of the form ${\cal M}\simeq \mathbb{R}\times M$ \cite{Geroch:1970uw} and one associates $\mathbb{R}$ with time and $M$ with space. Hence, the four-dimensional manifold ${\cal M}$ is considered as a foliation of three-dimensional space-like hypersurfaces $X_t(M)$ labelled by a parameter $t\in\mathbb{R}$, where $X_t:M\to{\cal M}$ is an embedding of the spatial manifold $M$ into ${\cal M}$. A particular choice of time and space would break diffeomorphism invariance and therefore in the framework of the ADM-formalism one does not choose a particular foliation but considers all possible ones. In the canonical framework the elementary configuration variables are the pull back of the metric $g_{\mu\nu}$ onto $M$ denoted by $q_{ab}$ from now on, also called the ADM 3-metric. The conjugate momenta, denoted by $p^{ab}$, are related to the extrinsic curvature of the space-like hypersurfaces $X_t(M)$. The diffeomorphism invariance of the theory has the consequence that general relativity is a constrained Hamiltonian theory meaning that in addition to the Hamiltonian equation of motion for $q_{ab}$ and $p^{ab}$ the theory possesses constraints, which are additional equations on phase space, that $q_{ab}$ and $p^{ab}$ have to satisfy. Therefore the constraints select out of the kinematical degrees of freedom $(q_{ab},p^{ab})$, which still include gauge degrees of freedom, the physical degrees of freedom. In the case of the ADM-formalism these constraints are called Hamiltonian and (spatial) diffeomorphism constraint. The latter generates diffeomorphisms within the spatial hypersurface $M$ and the Hamiltonian constraints are generating diffeomorphisms orthogonal to the hypersurface. Note that in the case of the Hamiltonian constraint this is only true on shell, that is when the constraints are satisfied, and in addition when the equation of motion are fulfilled. Furthermore, general relativity is in this sense special as its Hamiltonian consists entirely of a linear combination of the constraints and therefore general relativity is called a fully constrained theory. This property has to be taken into account when one discusses the notion of observables, that is gauge invariant quantities, in the context of general relativity, see for instance \cite{RRF,DObs}.
\\
\\
As far as the quantization of theories with constraints is considered, there exist two different approaches to formulate the quantum theory. One, known as Dirac quantization, goes back to Dirac \cite{Dirac} and is based on the idea to quantize the entire kinematical phase space including the gauge degrees of freedom yielding the so called kinematical Hilbert space ${\cal H}_{\rm kin}$. Let us denote the set of classical constraints by $\{{\cal C}_I\}$ with $I\in{\cal I}$ where ${\cal I}$ denotes some arbitrary index set. Then the physical sector of the theory is constructed in the quantum theory by implementing all classical constraints  $\{{\cal C}_I\}$ as operators $\{\hat{C}_I\}$ on ${\cal H}_{\rm kin}$ and requiring that physical states $\psi$ are those, which are annihilated by all constraints operators, that is $\hat{\cal C}_I\psi=0$ for all $I\in{\cal I}$. These physical states are elements of the so called physical Hilbert space ${\cal H}_{\rm phys}$.
The second approach called reduced quantization follows the strategy to solve the constraints already at the classical level. In doing so, one obtains the reduced or also called physical phase space whose elementary variables are called observables because there are gauge invariant quantities and do not include gauge degrees of freedom any longer. Then one quantizes the physical phase space, which corresponds to the task of finding suitable representations of the algebra of observables leading directly to the physical Hilbert space ${\cal H}_{\rm phys}$. In addition one is only interested in those representations which also allow to implement the dynamics of those observables in the quantum theory.
\\ 
Now in practice one often does not exclusively follow Dirac or reduced quantization but often combines both approaches. If for example the classical constraints are complicated to solve, Dirac quantization might be of advantage as long one is able to solve the corresponding quantum constraint equations. On the other hand, if one is able to reduce the constraints at the classical level, one quantizes only the physical phase space and has thus a direct access to the physical Hilbert space one is finally interested in in both approaches. The technical difficulty in the reduced quantization occurs when the resulting algebra of observables has a much more complicated structure than the corresponding kinematical one because it might be impossible to find representations of the algebra and hence to formulate the quantum theory at all.
\\
\\
As far as the ADM-variables are concerned one has mainly followed the Dirac quantization procedure and used standard Schr\"odinger quantization techniques known from other quantum field theories to construct the corresponding kinematical Hilbert space ${\cal H}_{\rm kin}$ for general relativity   \cite{DeWitt:1967ub}. If we denote the diffeomorphism constraint by ${\cal C}_{\rm Diff}$ and the Hamiltonian constraint by ${\cal C}$ then one needs to find quantum states $\psi(q_{ab})$ that satisfy $\hat{\cal C}_{\rm Diff}\psi(q_{ab})=0$ and $\hat{\cal C}\psi(q_{ab})=0$. The latter equation involving the Hamiltonian constraint is also known as the Wheeler--DeWitt equation. However, this quantization of general relativity has to be understood rather at a formal level because not all details about the measure underlying ${\cal H}_{\rm kin}$ have been worked out. Also, using ADM-variables and the standard Schr\"odinger representation it has up to now not been shown that the Hamiltonian constraint operator can be implemented on ${\cal H}_{\rm kin}$ as a well defined operator. Exactly these difficulties have been the starting point for reconsidering the canonical quantization of general relativity from a different angle. We will see in the next section that a different choice of elementary variables, called connection or Ashtekar variables, to describe the canonical formulation of general relativity will allow to formulate the kinematical Hilbert space ${\cal H}_{\rm kin}$ for general relativity not only at the formal level and will allow to implement \underline{all} constraints of general relativity as operators on ${\cal H}_{\rm kin}$.
\section{General Relativity in Connection Variables}
The motivation for deriving a formulation of general relativity  in terms of connection variables is that it allows to describe general relativity in a language very close the language that is used in other quantum field theories for which already powerful quantization techniques exist.  

The starting point for the connection formulation is to describe general relativity in terms of frames. A frame field denoted by $e_I:=e^\mu_I\partial_\mu$ with $I=0,1,2,3$ defines a basis of the tangent space $T_p{\cal M}$ at each point $p$ of ${\cal M}$. Here we will discuss the connection formulation by starting already with the ADM 3+1-split of the space-time and therefore work with frame fields, which are a point dependents basis for the tangent space $T_pM$ associated with the 3 dimensional manifold $M$. Usually one works with orthonormal frames, meaning that $e_j:=e^a_j\partial_a$ with $j=1,2,3$ satisfy
\begin{equation}
\langle e_j, e_k\rangle=e^a_je^b_k q_{ab}=\eta_{ij}, 
\end{equation}
where $\eta_{jk}$ denotes the components of the Euclidian metric on $\mathbb{R}^3$ and $e^a_j$ is called triad or 3-bein respectively. Given a frame field, we can define the (inverse) 3-metric in terms of the triads
\begin{equation}
\label{invmetrictriads}
q^{ab}=e^a_je^b_k\eta^{jk}.
\end{equation}
Conversely, $q^{ab}$ defines a triad, however only up to $SO(3)$-rotations. Likewise to a frame we can also introduce a co-frame field $e^j:=e^j_a dx^a$ being a point dependent basis for the co-tangent space $T_p^*M$. At each point in $M$ we can view $e_j^a$ and $e^j_a$ as  non-singular matrices. Using the isomorphism between the Lie-algebras of su(2) and so(3) we can regard $e^j_a$ as an su(2)-valued one-form. When we take this point of view we have to replace $\eta_{ij}$ by the Killing metric of su(2), which we will also denote by $\eta_{ij}$.
Since the co-frame is at each point the dual basis of the frame we have
\begin{equation}
\eta^j_k=e^j(e_k)=e_a^je^b_kdx^a(\partial_b)=e_a^je^b_k\delta_b^a=e_a^je^a_k.
\end{equation}
Furthermore, we have
\begin{equation}
\delta^a_b=q^{ac}q_{cb}=e^a_je^c_k\eta^{jk}e_c^me^b_n\delta_{mn}
=e^a_je^b_n\eta^{jk}\delta_k^m\delta_{mn}=e^a_je_b^j.
\end{equation}
Due to the additional $SO(3)$ freedom encoded in the triads the passage from the ADM-phase space to the frame formulation is not a canonical transformation but an extension of the ADM phase space. The elementary variable we will work with is not the triad itself but its densitized version, which is an su(2)-valued vector density of weight one, denoted by $E_j^a$ and defined as
\begin{equation}
E_j^a:=\sqrt{\det(q)}e^a_j,
\end{equation}
where $\det(q)=det(q^{-1})^{-1}=\det(e)^{-2}$ is understood as a function of the triads. The densitized triads will be the momentum variables in the new phase space. As mentioned before in the ADM phase space the canonically conjugate momenta to $q_{ab}$ are related to the extrinsic curvature, which we will denote by $K_{ab}$. The canonically conjugate configuration variable to the densitized triad is given by
\begin{equation}
K^j_a:=K_{ab}e^b_k\eta^{jk},
\end{equation}
which is, like $e^j_a$, a su(2)-valued one-form.
\\
\\
For the reason that we have extended the ADM phase space by additional rotational degrees of freedom encoded in the (co)-frames we obtain the so called rotational constraints given by
\begin{equation}
\label{RotConst}
G_j=\epsilon_{jk\ell}K^k_aE^a_\ell,
\end{equation}
which ensure, that on shell we obtain again the ADM phase space. Given the canonical pair $(K^j_a, E^a_j)$ we obtain the Ashtekar variables by applying two canonical transformations. The first one is a rescaling of the elementary variables, which introduces the so called Barbero--Immirzi parameter $\gamma\not=0\in\mathbb{C}$ into the classical theory
\begin{equation}
K^j_a\to{}^{(\gamma)}{K}^j_a:={\gamma}K^j_a\quad\quad
E^a_j\to{}^{(\gamma)}{E}^a_j:=\frac{1}{\gamma}E^a_j.
\end{equation}
The second canonical transformation involves the spin connection, which we briefly  discuss before describing the canonical transformation. Given the metric $q_{ab}$ on $M$ there exists a unique Levi-Cevita connection $\nabla$, also called covariant derivative, which is metric compatible, that is $\nabla q_{ab}=0$ and torsion-free, that is $\Gamma^a_{bc}=\Gamma^a_{cb}$, where $\Gamma^a_{bc}$ are the Christoffel symbols associated with $q_{ab}$. Once we introduce triads we have to consider tensors having spatial as well as su(2) indices and therefore we extend the covariant derivative onto tensors with mixed indices by defining
\begin{equation}
\nabla_a t^b_j:=\partial_a t^b_j+\Gamma^b_{ac}t^c_j+\Gamma^{\,\,k}_{a\,\,j}t^b_k, 
\end{equation}
with $\Gamma_{ajk}=-\Gamma_{akj}$ so that $\Gamma_a$ is an antisymmetric matrix and takes values in so(3).
The extension to arbitrary tensors is obtained by linearity, the Leibniz rule and the requirements that $\nabla_a$ commutes with contractions. If we extend the metric compatibility $\nabla_a q_{bc}=0$ to $\nabla_a e^b_j=0$ we can express $\Gamma^{\,\,k}_{a\,\,j}$ in terms of the (co)-triads and the Christoffel-symbols $\Gamma^a_{bc}$ given by
\begin{equation*}
\Gamma_{a\,\,k}^{\,\,j}=-e^b_k\left(\partial_a e^j_b-\Gamma^c_{ab}e^j_c\right).
\end{equation*}
Since $\Gamma_a$ takes values in so(3), we can use a basis of so(3) denoted by $\{T_1,T_2,T_3\}$ with $(T_i)_{jk}=\epsilon_{ikj}$ to expand $\Gamma_a$ as $\Gamma_a^j T_j$ with $\Gamma_a^j$ being the the spin connection. Note that we can also consider $T_i$ as the generators of su(2) in the adjoint representation since  there exists an isomorphism between su(2) in the adjoint and so(3) in the defining  representation. Using the spin connection we can now perform the second canonical transformation, which is an affine transformation, and finally leads to the connection or nowadays also called Ashtekar variables
\begin{equation}
{}^{(\gamma)}{K}^j_a\to {}^{(\gamma)}A^j_a:=\Gamma^j_a+{}^{(\gamma)}{K}^j_a\quad\quad
{}^{(\gamma)}{E}^a_j\to {}^{(\gamma)}{E}^a_j.
\end{equation}
Although $\Gamma^j_a$ has, as a function of $E^a_j$, a complicated form it was proven \cite{Ashtekar:1986yd, HNS1989}, that $({}^{(\gamma)}A^j_a,{}^{(\gamma)}{E}^a_j)$ build indeed a canonical pair and satisfy the following Poisson algebra
\begin{eqnarray}
\{{}^{(\gamma)}A^j_a(x),{}^{(\gamma)}A^k_b(y)\}&=&\{{}^{(\gamma)}{E}^a_j(x),{}^{(\gamma)}{E}^b_k(y)\}=0\\
\{{}^{(\gamma)}A^j_a(x),{}^{(\gamma)}{E}^b_k(y)\}&=&k\delta^j_k\delta_a^b\delta^3(x,y),
\end{eqnarray}
where $k=8\pi G_N$ with $G_N$ being Newton's constant. In order to absorb the factor $k$ occurring above in the definition of the elementary variables we use 
\begin{equation}
{}^{(\gamma)}P^a_j:=\frac{1}{k}{}^{(\gamma)}{E}^b_k(y)
\end{equation}
 as the canonically conjugate momentum to ${}^{(\gamma)}A^j_a(x)$ in the following.
 \\
Let us briefly comment on the role of the Barbero-Immirzi-parameter. For each choice of $\gamma$ we obtain a different set of canonical variables to coordinatize the phase space of general relativity. At this point the choice is arbitrary but might be determined from other physical situations like for instance the computation of the black hole entropy (see the chapter by Barbero and Perez in \cite{Book}). In the literature different choices of $\gamma$ have been discussed, as for example $\gamma=\pm i$ \cite{Ashtekar:1986yd} and $\gamma\in\mathbb{R}$ \cite{Barbero:1994ap} and $\gamma\in\mathbb{C}$ \cite{Immirzi:1996di}. The choice   $\gamma=\pm i$ is special in the sense that in this case the Hamiltonian constraint simplifies in the sense that i.) The Hamiltonian constraint --and consequently its later quantization-- simplifies and ii.) on classical solutions ${}^{(\gamma)}A^j_a$ has the natural geometric meaning of the restriction to $M$ of the self-dual part of the space-time Lorentz connection. However, in this case the connection ${}^{(\gamma)}A^j_a$ is complex leading to an additional reality condition for ${}^{(\gamma)}A^j_a$ whose implementation on the quantum level is highly non-trivial. Therefore currently, one mainly works with real $\gamma$ and real connection variables. From now on we will drop the label ${}^{(\gamma)}$ and just use $(A^j_a,P^a_j)$ in order to keep our notation more clearly and always keep in mind that the construction of the Ashtekar variables involves the Barbero-Immirzi-parameter.
\\
As mentioned in the last section the introduction of the Ashtekar variables allows to describe general relativity very close to the language of other gauge theories used in quantum field theory and this point will become clear when we discuss the form of the constraints in terms of Ashtekar variables. We saw that with the extension of the ADM phase space we obtained the rotational constraint in (\ref{RotConst}). Expressed in terms of $(A^j_a,P_j^a)$ it has the form
\begin{equation}
G_j=\partial_aE^a_j+\epsilon_{jk}^{\,\,\,\,\,\,\ell} A^k_aP^a_\ell=:{\cal D}_aP^a_j,
\end{equation}
where we introduced a new covariant derivative ${\cal D}$, that involves instead of the spin connection the SU(2) connection $A^j_a$. In terms of these new variables the rotational constraints have the form of an SU(2) Gauss law known from Yang-Mills gauge theory. Hence, in terms of the connection variables general relativity can be understood as a SU(2) gauge theory. The remaining constraint, that were already present in the ADM-formalism, are the (spatial) diffeomorphism constraint ${\cal C}_a$ and the Hamiltonian constraint ${\cal C}$. Using the connection variables and considering the $G_j=0$ constraint hypersurface, these are given by
\begin{equation}
{\cal C}_{a}=F^j_{ab}P^b_j\quad\quad
{\cal C}=\frac{k\gamma^2}{2}\frac{\epsilon_j^{\,\,\,mn}P^a_mP^b_n}{\sqrt{\det(q)}}\left(F^j_{ab}-(1+\gamma^2)\epsilon^{jk\ell}K^k_aK^m_b\right),
\end{equation}
where  we dropped the term proportional to the Gauss constraint in ${\cal C}_{a}$ and  $F^j_{ab}$ is the curvature associated with the connection $A^j_a$ 
\begin{equation}
F^j_{ab}=\partial_a A^j_b-\partial_b A^j_a+\epsilon^j_{\,\,k\ell}A^k_aA^\ell_b
\end{equation}
and $K^j_a=A^j_a-\Gamma^j_a$ is considered as a function of $(A,P)$ and $\det(q)$ as a function of $P$.
Let us introduce the smeared version of the above constraints
\begin{equation}
{\cal C}_G(\Lambda):=\int\limits_M d^3x (\Lambda^j G_j)(x),\quad
{\cal C}_{\rm Diff}(\vec{N}):=\int\limits_M d^3x (N^a {\cal C}_a)(x)\quad
{\cal C}(N):=\int\limits_M d^3x (N {\cal C})(x).
\end{equation}
Here $\Lambda^j$ is lie-algebra-valued smearing field and $N$ and $N^a$ are the lapse function and the shift vector respectively, which in the ADM-formalism are related to the $00$ and $0a$ components of the (inverse) metric $g^{\mu\nu}$ by
\begin{equation}
g^{00}=N^{-2},\quad g^{0a}=N^{-2}N^a.
\end{equation}
An aspect that will be later important when the quantization of the (smeared) constraints is discussed is that they satisfy the following constraint algebra
\begin{eqnarray}
\{{\cal C}_{G}(\Lambda),{\cal C}_{G}(\Lambda)\}&=&{\cal C}_G(\Lambda),\quad
\{{\cal C}_{G}(\Lambda),{\cal C}_{\rm Diff}(\vec{N})\}=-{\cal C}_G({\cal L}_{\vec{N}}\Lambda),\quad \\
\{{\cal C}_{\rm Diff}(\vec{N}),{\cal C}_{\rm Diff}(\vec{N}')\}&=&{\cal C}_{\rm Diff}({\cal L}_{\vec{N}}\vec{N}'),\quad
\{{\cal C}_{G}(\Lambda),{\cal C}(N)\}=0
 \\
\{{\cal C}(N),{\cal C}(N')\}&=&-{\cal C}_{\rm Diff}(\vec{S}),\quad
S^a:=\frac{P^a_jP^b_k\eta^{jk}}{|\det(q)|}\left(NN'_{,b}-N'N_{'b}\right).
\end{eqnarray}
The subalgebra of ${\cal C}(N)$ and ${\cal C}_{\rm Diff}(\vec{N})$ encodes the diffeomorphism invariance at the canonical level and can be also derived from purely geometrical considerations \cite{Hojman:1976vp}, see also the discussion in the chapter by Laddha and Varadarajan in \cite{Book}. It will play a pivotal role in the quantization of the constraint operators because one requires that the corresponding constraint operators satisfy an analogue commutator algebra in order to carry over the classical symmetries into the quantum theory.
\\ Let us finally summarize: We have formulated general relativity in terms of connection variables $(A,P)$. The corresponding action in the 3+1-picture is given by
\begin{equation}
S=\int\limits_{\mathbb{R}} dt\int\limits_M d^3x \left(\dot{A}^j_aP^j_a-\left(\Lambda^jG_j+N{\cal C}+N^a{\cal C}_a\right)\right).
\end{equation}
The 'Hamiltonian' $H$ is given by
\begin{equation}
H={\cal C}_G(\Lambda)+{\cal C}(N)+{\cal C}_{\rm Diff}(\vec{N})
\end{equation}
and, as mentioned before, is a linear combination of constraints only. The Hamiltonian equation of motion
\begin{equation}
\dot{A}^j_a(x)=\{A^j_a(x),H\},\quad\quad
\dot{P}_j^a(x)=\{P_j^a(x),H\}
\end{equation}
together with the constraints
\begin{equation}
{\cal C}_G(\Lambda)=0,\quad\quad{\cal C}(N)=0,\quad\quad
{\cal C}_{\rm Diff}(\vec{N})=0
\end{equation}
are completely equivalent to Einstein's equations in vacuum
\begin{equation}
R_{\mu\nu}-\frac{1}{2}g_{\mu\nu}R=0.
\end{equation}
For the reason that the 'Hamiltonian' $H$ vanishes on the constraint hypersurface, the evolution generated by $H$ is interpreted as gauge transformations and not as a physical evolution. A discussion how physical evolution can be implemented in the context of general relativity in the framework of observables can for instance be found in \cite{AQGIV}. 
\\
\\
We have discussed the connection formulation for space-times of dimension 4. In $D+1$ dimensions a spatial metric has $\frac{D(D+1)}{2}$ degrees of freedom, while a frame in $D$ dimensions includes $D^2$ degrees of freedom. Consequently, we need $D^2-\frac{(D(D-1)}{2}=\frac{D(D+1)}{2}$ constraints in order to recover the corresponding ADM formulation in $D+1$ dimensions. Note that $\frac{D(D+1)}{2}$ is precisely the dimension of $SO(D)$ and thus it would be a natural choice for a gauge group here. However, an $SO(D)$ connection has $\frac{D^2(D-1)}{2}$ degrees of freedom and the only dimension for $D$ in which the number of degrees of freedom of the $D$-bein and the $SO(D)$ connection coincide is the special case $D=3$.
 \\
However, this does not mean, that there exists no connection variable formulation in higher dimensions. Recently, it has been shown that one can introduce a different extension of the ADM phase space and formulate general relativity in terms of SO(D+1) Yang Mills variables \cite{TNv}. In order to match the degrees of freedom of the ADM phase space and thus general relativity, the formulation in \cite{TNv} includes additional constraints, that have to be implemented.
\\
\\
Although we have restricted our discussion to the vacuum case here, the connection formulation can be generalized to gravity coupled to matter by simply performing a 3+1-split also for the matter action \cite{ARTmatter} and see also \cite{Thiemann:2007zz} for a pedagogical introduction to this topic. We then obtain further degrees of freedom in phase space describing the matter part of the theory. The constraints and hence also the 'Hamiltonian' will then include additional contributions from the matter degrees of freedom. In the next section we will also restrict the discussion for simplicity to the vacuum case and show how a quantum theory for the connection formulation can be constructed.
\section{Holonomy Flux Algebra and its Representation(s)}
The connection formulation of general relativity  discussed in the last section is the classical starting point for loop quantum gravity. Before we explain in detail how this works for the connection formulation of general relativity let us briefly recall how canonical quantization is used in quantum mechanics. 
\subsection{Canonical Quantization in Quantum Mechanics}
In quantum mechanics we choose as the classical starting point the phase space coordinatized by $(q^j, p_j)$, that satisfy the so called Heisenberg algebra
\begin{equation}
\{q^j,q^k\}=0\quad\quad\{p_j,p_k\}=0\quad\quad \{q^j,p_k\}=\delta^j_k.
\end{equation}
To formulate the quantum theory, we introduce an abstract ${}^*$-algebra\footnote{A ${}^*$-algebra is an algebra with an involutions, that is a map $*:\mathfrak{A}\to\mathfrak{A}$ $a\mapsto a^*$ with the following properties $(ca+c'a')^*=\overline{c}a^*+\overline{c}'a'^*$, $(aa')^*=a'^*a^*$ and $(a^*)^*=a$ for all $a,a'\in\mathfrak{U}, c,c'\in\mathbb{R}$.} ${\mathfrak U}$ of operators generated by $\hat{q}$, $\hat{p}$ and $\mathbbm{1}_{\mathfrak A}$. Since we want to replace Poisson brackets by commutators in the quantum theory we set
\begin{equation}
[\hat{q}^j,\hat{p}^k]=:i\hbar\widehat{\{q^j,p_k\}}\quad\quad (\hat{q}^j)^*=\overline{q}^j\quad\quad
(\hat{p}_j)^*=\overline{p}_j,
\end{equation}
where the bar denotes complex conjugation. The task is now to find a representation of this abstract ${}^*$-algebra, that is a map $\pi: \mathfrak{A}\to{\cal L}({\cal H})$ from the algebra into the subalgebra of linear operators on a Hilbert space ${\cal H}$, which has the following properties
\begin{equation}
\pi(c\hat{a}+c'\hat{a}')=c\pi(\hat{a})+c'\pi(\hat{a}')\quad
\pi(\hat{a}\hat{a}')=\pi(\hat{a})\pi(\hat{a}')
\quad
\pi(a^*)=\pi(\hat{a})^\dagger,
\end{equation}
where $a$ is an element of the algebra generated by $\{q^j,p_j,\mathbbm{1}_{\mathfrak{A}}\}$, ${}^\dagger$ denotes the adjoint operation and $\pi$ furthermore has to satisfy
\begin{equation*}
[\pi(\hat{q}^j),\pi(\hat{p}^k)]=i\hbar\pi(\mathbbm{1}_{\mathfrak{A}})=i\hbar\mathbbm{1}_{\cal H}.
\end{equation*}
In the case of quantum mechanics (QM) the representation is well known and called the Schr\"odinger representation. The Hilbert space ${\cal H}$ is ${\cal H}=L_2(\mathbb{R}^3,d^3x)$ and the explicit form of the representation is
\begin{equation}
(\pi(\hat{q}^j)\psi)(x)=x^j\psi(x)\quad\quad
(\pi(\hat{p}_j)\psi)(x)=-i\hbar\frac{\partial\psi}{\partial x^j}(x).
\end{equation}
Hence, the configuration variables become multiplication and the momenta derivation operators. We realize that formulating the classical theory requires two main choices for any quantum theory. The first choice is the classical Poisson algebra, that we take as a starting point for the quantization. Different choices will in general lead to different algebras and therefore finally also to different quantum theories. Secondly, even if we restrict our discussion to one particular choice of the classical Poisson algebra, in general there exists more than one possible representation of this algebra. Any of those representations can in principle define a different quantum theory, unless they are unitary equivalent. We call two representations $\pi_1$ and $\pi_2$ unitary equivalent if there exists an unitary operator $U:{\cal H}_1\to{\cal H}_2$ such that $U\pi_1(\hat{a})U^{-1}=\pi_2(\hat{a})$ for all $\hat{a}\in\mathfrak{A}$. In the context of QM the famous Stone-von-Neumann uniqueness theorem states that under very weak assumptions on the representation the Schr\"odinger representation is up to unitary equivalence the unique representation for QM. This theorem was announced by Stone in 1930 and the first complete proof was given by von Neumann \cite{J.Neumann}. 
The actual proof uses the Weyl- instead of the Heisenberg-algebra, whose generators are the exponentiated versions of the $q^j$'s and $p_j$'s discussed above. However, since one of the assumptions for the representation is that it should be weakly continuous, the operators $\hat{q}^j$ and $\hat{p}_j$ also exists in this representation and one can also recover the Heisenberg commutation relations coming from the Weyl-algebra. So far we have only considered kinematical requirements for the choice of the representation. Of course the dynamics plays as an important role as it does in the classical theory. Therefore, we are only interested in those representations that allow to implement the generators of the classical dynamics as operators. In the case of standard QM, this is the Hamiltonian, which usually is a polynomial on phase space. Hence, in the Schr\"odinger representation the corresponding operators can be implemented. In the case of general relativity using Dirac quantization we have to find representations for which the classical constraints can be quantized on the kinematical Hilbert space. We will see in the following discussion, that this requirement forces us to introduce a different representation than the usual Fock representation used in standard quantum field theory.
\subsection{The Holonomy--Flux--Algebra $\mathfrak{A}$}
Now we take the connection formulation of general relativity as our classical starting point for the quantization. The difference with classical mechanics is that general relativity is a field theory and hence the variables $(A^j_a(x), P_j^a(x))$ are too singular to be directly promoted to operators. Therefore one quantizes not $(A^j_a(x), P_j^a(x))$ themselves but particular smeared versions of these elementary variables. In the case of standard canonical quantum field theory, one uses a 3 dimensional smearing over $M$ for the basic field variables and their conjugate momenta. However, this kind of smearing is defined with respect to a particular background metric. For general relativity we will choose a different way of smearing  $(A^j_a(x), P_j^a(x))$, which in particular has the property to be independent of any background metric and leads to basic variables similar to those used in ordinary lattice gauge theory. The SU(2)-connection $A^j_a$ is an su(2)-valued one-form and thus it is natural to integrate the connection along oriented curves $e:[0,1]\to M, s \mapsto e(s)$ in $M$, which we call edges. If we further take the path-ordered exponential of this integral, we obtain the holonomy associated with the connection $A$ given by
\begin{eqnarray}
A(e)&:=&{\cal P}\exp(\int\limits_e A)\\
&=&\mathbbm{1}_2+\sum\limits_{n=0}^\infty\int\limits_0^1 ds_1\int\limits_{s_1}^1 ds_2\cdots\int\limits_{s_{n-1}}^1 ds_n A(e(s_1))
A(e(s_2))\cdots A(e(s_n)),
\end{eqnarray}
where $A(e(s_i)):=A^j_a(e(s_i))\tau_je^a(s_i)$ and $\tau_j$ denotes a basis of su(2). 
Let us consider an edge $e:[0,1]\to M, s\mapsto e(s)$ in $M$ with beginning point $b(e)=e(0)$ and final point $f(e)=e(1)$ and let $t\in[0,1]$. Then the holonomy $A(e)=A(e,1)$ is the unique solution of the following differential equation
\begin{equation}
\frac{d}{dt}A(e,t)=A(e,t)A^j_a(e(t))\tau_j\dot{e}^a(t)\quad{\rm with}\quad A(e,0)=\mathbbm{1}_2
\end{equation}
which describes the parallel transport from  $b(e)$ to $f(e)$ along the edge $e$.
In our case the holonomy is an element of the group SU(2). Under the composition of two edges $e_1\circ e_2$, for which the final and beginning point are the same and under the inversion of edges $e^{-1}$, the holonomy behaves as
\begin{equation}
A(e_1\circ e_2)=A(e_1)A(e_2)\quad\quad A(e^{-1})=A^{-1}(e).
\end{equation}
Note that $e^{-1}$ is obtained from $e$ by reversing the orientation of the edge.
\\
Similar variables are also used in ordinary lattice gauge theory with the corresponding connections of the gauge theories of the standard model. The reason for this is that the holonomies transform very simply under gauge transformations. While the connection transforms as $A^g=gAg^{-1}-dgg^{-1}$  under SU(2) gauge transformation, the transformed holonomy is $A^g(e)=g(e(0))A(e)g(e(1))^{-1}=g(b(e))A(e)g(f(e))^{-1}$. Hence, the transformation acts only at the beginning and final points of the curve and this simple behavior is of advantage when later gauge invariant quantities in the quantum theory will be constructed. For instance the famous Wilson-loop defined as $Tr({\cal P}\exp(\oint\limits_\beta A))$  is the holonomy of a given connection $A$ along a closed loop $\beta$ and one example of a gauge invariant observable because the trace allows to cyclic permute the matrices and $b(e)=f(e)$ for a loop  so that $g(f(e))^{-1}g(b(e))=\mathbbm{1}_G$ can be used, where $\mathbbm{1}_G$ denotes the unit element in the gauge group $G$. Considering the conjugate variable $P_j^a$ also here exists a --from the geometric perspective-- natural smearing. The densitized triad $P_j^a$ is a su(2)-valued vector density of weight +1. Introducing a su(2)-valued smearing field $f^j$, $P^a(f):=f^jP_j^a$ is vector density and hence dual to a (pseudo-) 2-form in three dimensions, using that $\epsilon_{abc}$ carries density weight -1. Given this (pseudo-) 2-form the natural smearing is over two-dimensional surfaces, thus we define the conjugate variables, the so called (electric) fluxes\footnote{The name (electric) flux is due to the fact that in the canonical version of electrodynamics the canonical momentum is precisely the electric fields and integrating it over a surface gives the electric flux.} as
\begin{equation}
\label{DefP(S,f)}
P(S,f)=\int\limits_S f^j(*P)_j=\int\limits_S f^j\epsilon_{abc}P^a_jdx^b\wedge dx^c.
\end{equation}
If one computes the Poisson bracket between the holonomies and fluxes the result depends on the position of the edge $e$ relative to the surface $S$. In order to discuss this in detail we introduce the notion of an elementary edge. We have to consider 4 different cases for the elementary edges. If $S\cap e=0$ we call and edge of type out. If $S\cap e=e$  and  hence $e$ lies entirely inside $S$ we call $e$ of type in. If $e$ is not of type in but $S\cap e\not=0$ we consider as elementary edges only those, which have one intersection point, denoted by $p$, with $S$ in its end points. If $e$ lies above $S$, we call $e$ of type up and if $e$ lies below $S$ of type down. Furthermore, we distinguish the cases where $p$ is the beginning point $b(e)$ and the final point $f(e)$ respectively. Any edge $e$ can be written as a composition of elementary edges by introducing appropriate additional vertices. Using this classification we have
\begin{equation}
\{A(e),P(S,f)\}=-\frac{\kappa(S,e)}{2}\times\begin{cases} A(e)\tau_j f^j(b(e)) &{\rm if}\quad S\cap e=b(e) \\
-\tau_j f^j(f(e))A(e) & {\rm if}\quad S\cap e=f(e), 
\end{cases}
\end{equation}
with
\begin{equation}
\kappa(S,e)=\begin{cases} +1 & {\rm if}\,\, $e$\,\, {\rm of\,\, type\,\, up} \\
 \,\,\,\,0 & {\rm if}\,\, $e$\,\, {\rm of\,\, type\,\, in\,\,or\,\,out} \\
  -1 & {\rm if}\,\, $e$\,\, {\rm of\,\, type\,\, down}. 
\end{cases}
\end{equation}
As discussed in the context of QM, we have to find a suitable Poisson algebra, which encodes the underlying classical theory. In the case of QM this was the Heisenberg- and Weyl-algebra respectively. In both cases the Hilbert space associated with the representation is an $L_2(\mathbb{R}^3,d^3x)$-space, that is the space of square integrable function over $\mathbb{R}^3$ with the standard Lesbegue measure $d^3x$ on $\mathbb{R}^3$. Hence, we see for QM the Hilbert space underlying the representation involves the construction of a measure on $\mathbb{R}^3$, which is the classical configuration space for classical mechanics. For general relativity we consider a classical field theory and in terms of the connection formulation the classical configuration space ${\cal A}$ is the space of smooth connections. As usual in canonical field theories, the quantum theory is not based on the classical configuration space, but requires the introduction of a larger space, that includes not only smooth connections but also so called generalized or distributional connections and is called the quantum configuration space denoted by $\overline{\cal A}$. Thus, for loop quantum gravity, we have to construct a measure on the quantum configuration space. For this reason we will choose our classical Poisson algebra in such a way that it can be easily extended from the classical configuration space ${\cal A}$ to the quantum configuration space $\overline{\cal A}$. For this purpose, we introduce so called cylindrical functions on ${\cal A}$. 
So far we have restricted our discussion to an arbitrary but single edge $e$. Now we generalize this picture and introduce the notion of a graph $\alpha$. A graph $\alpha$ consists of a finite collection of edges $\{e_1,\cdots, e_n\}$ in $M$, whereas the edges intersect only in their beginning or final points. This intersection points are called vertices of  $\alpha$. For a given graph $\alpha$, we denote the set of edges by $E(\alpha)$ and the set of vertices by $V(\alpha)$. In order to give the definition of a cylindrical function, we denote the subset of connections associated with a graph  $\alpha$ by ${\cal A}_\alpha\subset{\cal A}$. ${\cal A}_\alpha$ contains all connections $A_{e_i}$ associated with the edges $\{e_i\}$ of the graph $\alpha$.  
Then there exists a map 
\begin{equation}
\label{IEmap}
I_E:{\cal A}_\alpha\to SU(2)^n\quad{\rm with}\quad A\in{\cal A}_\alpha\mapsto I_E(A):=(A(e_1),\cdots,A(e_n))
\end{equation}
 and we can use the map $I_E$ to define smooth cylindrical functions\footnote{Here a cylindrical function $f$  is said to be smooth if any of its representatives $f_\alpha$ on $G^n$ is smooth.} 
 defined with respect to a given graph $\alpha$ with edges $\{e_1,\cdots, e_n\}$ as
\begin{equation}
f_\alpha(A)=F_\alpha(I_E(A))=F_\alpha(A(e_1),\cdots,A(e_n)), 
\end{equation}
where $F_\alpha:SU(2)^n\to \mathbb{C}$ is a $C^\infty-$function on n copies of SU(2). A  function $f$ on ${\cal A}$ is said to be cylindrical if it can be written in the above form for some graph $\alpha$. Since each $f_\alpha$ depends only on a finite number of holonomies, we need to consider all possible graphs $\alpha$, that can be embedded into $M$ in order to describe the Poisson algebra underlying gravity in connection variables. A graph $\alpha'$ is said to be larger than a given graph $\alpha$, if every edge $e$ can be written as a finite combination of edges $e'_i$ of $\alpha'$, that is $e=e^{\prime s_1}_1\circ\cdots\circ e^{\prime s_\ell}_\ell$ for some set of edges $\{e'_i\,|\, i=1,\cdots,\ell\}$ of $\alpha'$ where $s=\pm 1$. Note that every function $f$ on ${\cal A}$, which is cylindrical with respect to a given graph $\alpha$ will automatically be cylindrical with respect to any larger graph $\alpha'$. This allows to define an equivalence relation on $\bigcup\limits_\alpha Cyl_\alpha$. Given $f,f'\in\bigcup\limits_\alpha Cyl_\alpha$ we can find $\alpha,\alpha'$ such that $f\in Cyl_\alpha$ and $f'\in Cyl_{\alpha'}$. We say that $f$ and $f'$ are equivalent, denoted by $f\sim f'$, provided that $f,f'$ agree for all larger graphs $\alpha''>\alpha,\alpha'$. We define the space of smooth cylindrical functions on ${\cal A}$ as
\begin{equation}
Cyl:=\bigcup\limits_\alpha Cyl_\alpha/\sim.
\end{equation}
Thus, $Cyl$ consists of equivalence classes of functions on the spaces $Cyl_\alpha$. $Cyl$ can be shown to be an Abelian $C^*$-algebra defined by pointwise operations and with the supremum-norm. In order to choose the Poisson algebra underlying loop quantum gravity, we still have to discuss the conjugate momentum variables associated with the smooth cylindrical functions on ${\cal A}$. The latter will be the flux vector fields on $Cyl$, which we  denote by $X(f,S)\in V(Cyl)$ and which are the Hamiltonian vector fields of $P(S,f)$, where $V(Cyl)$ includes not only the Hamiltonian but all vector fields on $Cyl$.
 The action of $X(S,f)$ on $f_\alpha$ is given by
\begin{flalign}
(X(S,f)f_\alpha)(A)&:=(\{f_\alpha,P(S,f)\})(A)&\\
=&\frac{k}{2}
\sum\limits_{e\in E(\alpha)} \frac{\kappa(e,S)}{2}\begin{cases}
A(e)\tau_j f^j(b(e)) &{\rm if}\quad S\cap e=b(e)\\
-\tau_j f^j(f(e))A(e) & {\rm if}\quad S\cap e=f(e)\end{cases}
\frac{\partial F_\alpha(\{A(e)_{e\in E(\alpha)}\})}{\partial A(e)_{AB}}& \nonumber
\end{flalign}
where $A,B$ denote SU(2)-indices. Finally, we can now define the classical Poisson algebra, which will be the starting point for our quantization in the next section, and which is called the holonomy--flux algebra $\mathfrak{A}$:
\begin{itemize}
\item
The classical Poisson algebra underlying loop quantum gravity is the Lie ${}^*$-subalgebra of $Cyl\times V(Cyl)$ generated by the smooth cylindrical functions and flux vector fields on $Cyl$. The involution on the algebra is just complex conjugation. This algebra is called the holonomy--flux algebra and will be denoted by $\mathfrak{A}$.
\end{itemize}
We will discuss the representation of the holonomy--flux algebra in the next section.
\section{The Ashtekar--Lewandowski Representation  and the kinematical Hilbert space of LQG}
So far we have discussed smooth cylindrical functions on the classical configuration space ${\cal A}$. For the derivation of the kinematical  Hilbert space underlying the representation of the holonomy--flux algebra, we have to construct a measure on the quantum configuration space $\overline{\cal A}$. The necessity of $\overline{\cal A}$ can be also understood from the following perspective: In order to obtain a kinematical Hilbert space ${\cal H}$ from $Cyl$, we need to take the Cauchy-completion with respect to a norm defined on $Cyl$. This completion will include objects as limit points, which cannot be understood as functions on ${\cal A}$, but are more general objects such as distributions on ${\cal A}$. The strategy one adopts is to look for a larger quantum configuration space $\overline{\cal A}$ such that ${\cal H}$ is isomorphic to an $L_2$-space over $\overline{\cal A}$ with respect to some measure on $\overline{\cal A}$. As we will see below the action of the flux vector fields on $Cyl$ can be easily extended from cylindrical functions on ${\cal A}$ to cylindrical functions on $\overline{\cal A}$ by the introduction of left- and right-invariant vector fields on SU(2). 
 A measure on $\overline{\cal A}$ can be defined by using the fact that any cylindrical function over a graph $\alpha$ can be expressed via the map $I_E$ in (\ref{IEmap}) by means of functions $F$ on ${\rm SU(2)}^n$. On $SU(2)^n$ a natural measure exists, using n copies of the Haar measure on SU(2). This allows to firstly define a measure on $\overline{\cal A}_\alpha$, which includes all, not necessarily smooth connections $\{A_{e_i}\}$ along the edges of the graph $\alpha$, and thus an inner product on $Cyl_\alpha$ for all $\alpha$ given by
 \begin{equation}
\langle f_\alpha, \tilde{f}_\alpha\rangle:=\int\limits_{{\rm SU(2)}^n} \prod\limits_{i=1}^nd\mu_H(A(e_i))\overline{F_\alpha(A(e_1),\cdots, A(e_n))}F_\alpha(A(e_1),\cdots, A(e_n)), 
 \end{equation}
 where $d\mu_H(g)$ denotes the Haar measure on SU(2). Taking the closure of $Cyl_\alpha$ with respect to the corresponding norm of the above defined inner product, we obtain Hilbert spaces ${\cal H}_\alpha:=L_2(\overline{A}_\alpha,d\mu_\alpha)$ for all graphs $\alpha$. The kinematical Hilbert space ${\cal H}$ can then be constructed using projective techniques, because $\overline{\cal A}$ can be understood as the projective limit of the $\overline{A}_\alpha$'s. Given the measures $\mu_\alpha$ on $\overline{A}_\alpha$, they can be used to construct a measure denoted by $\mu_{AL}$, called the Ashtekar-Lewandowski measure,  on $\overline{\cal A}$. For this purpose, we have to discuss how an inner product can be defined in case the functions $f_{\alpha'},\tilde{f}_{\alpha''}$ are cylindrical with respect to two different graphs $\alpha'$ and $\alpha''$ respectively.
Given this situation, we can use that $Cyl$ has the property, that we can always find a common graph $\alpha>\alpha',\alpha''$ with respect to which $f,\tilde{f}$ are cylindrical. Hence, we can use $\alpha$ to define an inner product for $f_{\alpha'},\tilde{f}_{\alpha''}$. Here we associate trivial holonomies to $f_{\alpha'}$ and $f_{\alpha''}$ respectively to those edges in $\alpha$, which are not contained in $\alpha'$ and $\alpha''$ respectively. 
Cylindrical consistency ensures that the inner product on $\overline{\cal A}$ does not depend on the particular choice of the common graph $\alpha$. For instance, if we take as the common graph just the union $\alpha:=\alpha'\cup\alpha''$, then the inner product defined with respect to $\alpha$ should yield the same value as if the we further unify the graph $\alpha$ with another graph $\alpha'''$ not contained in $\alpha'$. Also, the inner product should be the same for two graphs $\alpha$ and $\tilde{\alpha}$ when $\tilde{\alpha}$ can be obtained from $\alpha$ just by subdividing edges of $\alpha$ by means of the introduction of additional vertices. Thus, we define the inner product on $\overline{\cal A}$ for $f,\tilde{f}\in Cyl$ as
\begin{equation}
\langle f_\alpha, \tilde{f}_\alpha\rangle:=\int\limits_{{\rm SU(2)}^n} \prod\limits_{i=1}^nd\mu_H(A(e_i))\overline{F_\alpha(A(e_1),\cdots, A(e_n))}F_\alpha(A(e_1),\cdots, A(e_n)),  
 \end{equation}
 where $\alpha$ is a common graph with respect to which $f$ and $\tilde{f}$ are cylindrical. Considering the closure of $Cyl$ with respect to the corresponding norm gives the kinematical Hilbert space ${\cal H}=L_2(\overline{\cal A},d\mu_{AL})$, which is the space of square integrable functions over $\overline{\cal A}$ with respect to the Ashtekar-Lewandowski measure. Now, given the kinematical Hilbert space ${\cal H}$ we can discuss the representation $\pi$ of the holonomy--flux algebra. The space $Cyl$ is dense in ${\cal H}$ and therefore we can define the action of the elementary operators in the Ashtekar-Lewandowski representation on $Cyl$.
 The holonomy operators act as multiplication operators and hence we obtain for cylindrical functions 
 \begin{equation}
 (\pi(f)\psi)(A)=(\hat{f}\psi)(A)=f(A)\psi(A)
 \end{equation}
 for $\psi\in{\cal H}$.
The flux vector fields become derivation operators and their explicit action is given by
\begin{equation}
 (\pi(P(S,f))\psi)(A)=\hat{P}(S,f)\psi(A)=(X(S,f)\psi)(A)
\end{equation}
for $\psi\in{\cal H}$, that lie in the domain of $\hat{P}(S,f)$. We will express the righthand side of the, equation above now by means of the left- and right-invariant vector fields on SU(2) denoted by $L_j$ and $R_j$ respectively. Given a function $f:SU(2)\to\mathbb{C}$ and $g\in$SU(2) these are defined as
\begin{equation}
(R_j f)(g):=\frac{d}{dt}\left(f(e^{t\tau_j}g)\right)_{t=0}\quad\quad
(L_j f)(g):=\frac{d}{dt}\left(f(ge^{t\tau_j})\right)_{t=0}.
\end{equation}
Thus, we can define the action of the flux operators on $f_\alpha$ in $Cyl_\alpha$ as
\begin{equation}
\label{FluxAction}
\hat{P}(S,f)f_\alpha(A)=\frac{\hbar}{2}\sum\limits_{v\in V(\alpha)} f^j(v)\sum\limits_{\substack{e\in E(\alpha)\\ e\cap v\not=\emptyset}} \kappa(e,S)\hat{Y}_j^{(v,e)}f_\alpha(A), 
\end{equation}
with
\begin{equation}
\label{Yoperator}
\hat{Y}_j^{(v,e)}:=\mathbbm{1}_{\cal H}\times\mathbbm{1}_{\cal H}\times\cdots\times\mathbbm{1}_{\cal H}\times\left\{\begin{array}{l} \,\,\,\,iR_j^{e} \\ -iL_j^{e} \end{array}\right\} \times\mathbbm{1}_{\cal H}\times\cdots\times\mathbbm{1}_{\cal H},\quad{\rm if}\quad\left\{\begin{array}{l}e\,\,{\rm outgoing}\,\,{\rm at}\,\, v \\ e\,\,{\rm ingoing}\,\,{\rm at}\,\, v 
\end{array}\right\}
\end{equation}
This finishes our discussion on the kinematical representation of loop quantum gravity. The next subsection will briefly deal with the question whether there exists other than the already introduced representation for the kinematical Hilbert space of loop quantum gravity.
\subsection{Other Representations of the Holonomy--flux-algebra $\mathfrak{U}$}
In the last section we discussed in detail  how the kinematical representation for loop quantum gravity looks like. As we have seen the algebra underlying loop quantum gravity is the holonomy--flux algebra $\mathfrak{U}$ and one possible representation of this algebra is the Ashtekar-Lewandowski representation (AL-representation) introduced above. In the context of quantum mechanics we already briefly mentioned that given a choice of a classical algebra in general more than one possible representation of the algebra exists and thus in general different quantum theories can be obtained from the same classical starting point. This is a particularly interesting aspect in the case of general relativity since it is in contrast to quantum mechanics a field theory and for those no Stone-Von Neumann theorem exists. As a consequence, in the context of field theories, in principle, infinitely many unitarily non-equivalent representations could exist. However, in practice finding representations of a given algebra can be a challenging task and often we are happy to have found one at all. Nevertheless it is an interesting question to ask what kind of assumptions in the AL--representation have to be required in order to make it, under those assumptions, the --up to unitary equivalence-- unique representation of the holonomy--flux algebra. 
\\
\\
An answer to this question is given by the so called LOST-theorem \cite{Lewandowski:2005jk,Fleischhack:2004jc} and yields progress in two directions. On the one hand, we learn what kind of characteristic properties the AL-representations has and on the other hand, we can try to look for new representations by violating one of those assumptions. What are the assumptions needed in the LOST-theorem? As required in most physical theories one of the assumptions is that the representation should be irreducible. This means that any vector in ${\cal H}$ is a cyclic vector. A cyclic vector $\Omega$ is a vector in ${\cal H}$ for which the set $\{\pi(a)\Omega \,|\, a\in\mathfrak{A}\}$ is dense in ${\cal H}$. The further assumptions are related to the (gauge) symmetries of general relativity formulated in connection variables. As usual for quantum theories one requires that the classical symmetries should be implemented by unitary operators. In the context of the holonomy--flux algebra $\mathfrak{U}$ the LOST theorem includes an assumption on a positive linear functional on the holonomy--flux algebra  so that this is automatically fulfilled  for the spatial diffeomorphisms and the SU(2)-gauge transformations. The positive linear functional is used in the context of the Gelfand--Naimark--Segal theorem to construct a cyclic representation of $\mathfrak{U}$. Moreover, the LOST-theorem assumes that there is at least one vector $\Omega\in{\cal H}$ that is invariant under diffeomorphisms. These assumptions are strong enough to restrict the number of possible representations of the holnomy-flux algebra, up to unitarily equivalence, to one single representations, the AL-representation, which is summarized in the theorem below \cite{Lewandowski:2005jk,Fleischhack:2004jc}
\begin{theorem}
There is only one cyclic representation of the holonomy--flux algebra $\mathfrak{A}$ with diffeomorphism
invariant cyclic vector - the Ashtekar-Lewandowski representation.
\end{theorem}
Characteristic properties of the AL-representation are:
\begin{itemize}
\item
As we will see in section \ref{Sec:GeomOp}  so called geometric operators associated with length, volume and area have purely discrete spectra, giving already an idea that quantum geometry could yield to a new fundamental picture of geometry.
\item 
Although operators for the holonomy $A(e)$ exist, there are no operators representing the connection $A^j_a$ directly in this representation. 
\item
Similarly, also for the spatial diffeomorphisms the infinitesimal generators do not exist, but only finite diffeomorphisms are implemented as unitary operators. 
\end{itemize}
A different representation, that is not unitary equivalent to the AL-representation was rather recently discussed in the literature and is the so called Koslowski-Sahlmann representation (KS--representation) \cite{Koslowski:2007kh, Sahlmann:2010hn, Koslowski:2011vn}. The way the LOST-theorem is circumvented is that in the KS--representation the spatial diffeomorphism are not implemented unitarily, as will be discussed more in detail below. In the context of the above mentioned GNS theorem, associated with the AL-representation is a so called GNS vacuum state, which in the case of AL-representation describes an extremely degenerate situation of an empty geometry. Here the smooth classical spatial geometry is expected to arise through some coarse-graining procedure that describes the transition from the deep quantum to the classical regime. Therefore an interesting question is whether the observed smoothness of classical geometry can already be described at the quantum level without applying any coarse-graining. Following this idea Koslowski \cite{Koslowski:2007kh} considered a slight modification of the AL-representation, in which he extended the representation of the flux operators. In particularly, the representation of the fluxes is changed by adding a c-number term
\begin{equation}
\pi_{P^{(0)}}(P(S,f))=\hat{P}(S,f)+P^{(0)}(S,f)\mathbf{1}_{\cal H}\quad\quad 
{\rm with}\quad\quad P^{(0)}(S,f):=\int\limits_S f^j(*P^{(0)})_j, 
\end{equation}
where $P^{(0)}(S,f)$ is the classical value of the flux with respect to a background geometry given by the densitized triad $E^{(0)}= k P^{(0)}$. For this reason we labeled the representation $\pi_{P^{(0)}}$ by $P^{(0)}$ in order to distinguish between the AL-- und KS--representation. The Hilbert space associated with $\pi_{P^{(0)}}$ is the same as in the AL-representation, that is ${\cal H}_{P^{(0)}}={\cal H}$ and the action of the holonomies and the cylindrical function respectively agrees, thus 
\begin{equation}
\pi_{P^{(0)}}(f)=\pi(f)=\hat{f}.
\end{equation}
Note that in the case $P^{(0)}=0$ we recover the AL-representation. In this sense the representations $\pi_{P^{(0)}}$ can be understood as a family of representations, with one member being the AL-representation. However, for other choices than $P^{(0)}=0$ the AL- and the KS--representation are not unitarily equivalent and therefore could in principle describe different physics. We have already mentioned above that spatial diffeomorphism are not implemented unitarily in the KS--representation, being however one of the assumptions in the LOST-theorem. By this we mean, that if $\hat{U}(\phi)$ denotes the unitary operator implementing spatial diffeomorphisms $\phi$ in the AL-representation, then in general we have
\begin{equation}
\hat{U}(\phi)\pi_{P^{(0)}}(P(S,f))\hat{U}^\dagger(\phi)\not= \pi_{P^{(0)}}(P(\phi(S),\phi^* f)).
\end{equation}
The reason that the above equality fails is that the quantity $P^{(0)}$ is fixed and will not transform under the action of $\hat{U}(\phi)$. Since also in the context of the KS--representation spatial diffeomorphism play an important role, it was shown in \cite{Sahlmann:2010hn} that by enlarging the Hilbert space ${\cal H}_{P^{(0)}}$ one can define unitary operators that implement spatial diffeomorphisms that also involve the background  field $P^{(0)}$ and hence implement the corresponding automorpisms of the SU(2) principle fibre bundle on which the whole mathematical formulation of the theory is based. In the context of the (enlarged) Hilbert space of the KS--representation, denoted by ${\cal H}_{KS}$, has an orthonormal basis of the form $\{|s,P^{(0)}\rangle\}$ where $s$ denotes a standard spin network in the AL-representation, which are discussed in detail in the next subsection and which provide an orthonormal basis for ${\cal H}$, and $P^{(0)}$ denotes, as before, a background field. The inner product in ${\cal H}_{KS}$ is of the following form
\begin{equation}
\langle s', P'^{(0)}\, | \, s, P^{(0)}\rangle=\langle s' \,|\, s\rangle_{\rm AL}\delta_{P'^{(0)},P^{(0)}}, 
\end{equation}
where $\langle s' \,|\, s\rangle_{\rm AL}$ denotes the inner product in the AL-representation. The action of the cylindrical functions and fluxes in ${\cal H}_{KS}$ is given by
\begin{equation}
\hat{f}|s,E\rangle=|\hat{f}s, E\rangle\quad\quad 
\hat{P}(S,f)|s,E\rangle=|\hat{P}(S,f)s,E\rangle + P^{(0)}(S,f)|s,E\rangle.
\end{equation}
As shown in \cite{Sahlmann:2010hn,Campiglia:2013nva} the KS--representation based on this enlarged Hilbert space supports a unitary implementation of the spatial diffeomorphisms as well as the SU(2) gauge transformation (for which similar problems occur) and the diffeomorphism and SU(2) gauge invariant Hilbert space can be constructed using the technique of group averaging that is discussed more in detail in the chapter by Laddha and Varadarajan in \cite{Book}. 
\\
In \cite{Stottmeister:2013qra} it was pointed out that if one considers also higher order commutators such as for instance the element of $\mathfrak{U}$ given by $(0,[\hat{P}(S,f_1),[\hat{P}(S,f_2),\hat{P}(S,f_3))]])$ then one can derive the following identity for the double commutator
\begin{equation}
[\hat{P}(S,f_1),[\hat{P}(S,f_2),\hat{P}(S,f_3))]]=\frac{1}{4}\hat{P}(S,[f_1,[f_2,f_3]])
\end{equation}
and thus using the AL-representation of the holonomy--flux algebra $\mathfrak{U}$ the elements $(0,[\hat{P}(S,f_1),[\hat{P}(S,f_2),\hat{P}(S,f_3))]])$ and $(0,\frac{1}{4}\hat{P}(S,[f_1,[f_2,f_3]])$ need to be identified.
Let us now consider the situation in the KS--representation. There we have 
\begin{eqnarray}
[\pi_{P^{(0)}}(P(S,f_1)),[\pi_{P^{(0)}}(P(S,f_2)),\pi_{P^{(0)}}(P)(S,f_3))]]\\
&=&[\hat{P}(S,f_1),[\hat{P}(S,f_2),\hat{P}(S,f_3))]],\nonumber
\end{eqnarray}
where the equality above is true because the constant contributions of the background fields $P^{(0)}$ cancel in the double commutator. As a consequence we obtain
\begin{equation}
[\pi_{P^{(0)}}(P(S,f_1)),[\pi_{P^{(0)}}(P(S,f_2)),\pi_{P^{(0)}}(P)(S,f_3))]]=\frac{1}{4}\hat{P}(S,[f_1,[f_2,f_3]]).
\end{equation}
However, due to the part coming from the background field in $\pi_{P^{(0)}}(P(S,f))$ we have 
\begin{equation}
\pi_{P^{(0)}}(P(S,[f_1,[f_2,f_3]]))\not=\hat{P}(S,[f_1,[f_2,f_3]]).
\end{equation}
The suggestion in \cite{Stottmeister:2013qra} to cure this problem is the modification of commutation relations of the standard holonomy--flux algebra by an appropriate central term. As also discussed in  \cite{Stottmeister:2013qra} it is still an open question whether the introduction of such a central term is sufficient in the context of further higher order commutators, that could yield additional relations among the algebra elements.
\\ 
 A different point of view is taken in \cite{Campiglia:2014hoa} where the holonomy--flux algebra is extended by the so called background exponentials denoted by $\beta_{P^{(0)}}(A)$ whose explicit form is given by
\begin{equation}
\beta_{P^{(0)}}(A):=e^{i\int\limits_S P^{(0)}\cdot A}\quad\quad {\rm with}\quad\quad P^{(0)}\cdot A:=(P^{(0)})^a_i A^i_a.
\end{equation}
Next to the holonomy and flux action in ${\cal H}_{KS}$ given above these background exponentials act as
\begin{equation}
\hat{\beta}_{P'^{(0)}}|s,P^{(0)}\rangle = |s,P'^{(0)}+P^{(0)}\rangle.
\end{equation}
We have discussed in the last section that the AL-representation is a representation of the holonomy--flux algebra $\mathfrak{U}$. If we instead consider the holonomy--flux algebra enlarged by these background exponentials, called the holonomy-background-exponential-flux algebra in \cite{Campiglia:2014hoa}, then it was shown in \cite{Campiglia:2014hoa} that the KS--representation can be also understood as a representation of the holonomy-background-exponential-flux algebra.
\subsection{Spin Networks as an Orthonormal Basis of the Kinematical Hilbert Space}
A useful orthonormal basis of the kinematical Hilbert space ${\cal H}$ is given by so called spin network basis. Also here we will take advantage of an already existing natural orthonormal basis in the Hilbert space $L_2(SU(2),d\mu_H)$.  Let us consider the equivalence classes of finite dimensional, unitary, irreducible representations of SU(2) on a representation space $V_j$ and take one representative of it denoted by $\pi^j$. We denote the dimension of $\pi^j$ by ${\rm dim}(\pi^j)$. We define the following functions on SU(2)
\begin{equation}
b^j_{mn}: {\rm SU(2)}\to\mathbb{C},\quad g\mapsto \langle g\, |\, b^j_{mn}\rangle:=\sqrt{{\rm dim}(\pi^j)}\pi^j_{mn}(g),\quad m,n=1,\cdots,{\rm dim}(\pi^j).
\end{equation}
Using the Haar measure $\mu_H$ on SU(2), we can define an inner product for $b^j_{mn}$ as
\begin{equation}
\langle b^j_{mn},b^{j'}_{m'n'}\rangle:=\int\limits_{\rm SU(2)} d\mu_H(g)\sqrt{2j+1}\pi^j_{mn}(g)\sqrt{2j'+1}\pi^{j'}_{m'n'}(g),
\end{equation}
where we used that ${\rm dim}(\pi^j)=2j+1$ in the case of SU(2). 
The Peter--Weyl theorem  proves that the set of functions $\{b^j_{mn}\}$ build an orthonormal basis of $L_2(SU(2),d\mu_H)$. In particular the proof is true for any compact Lie group $G$. Hence, in our case $G=$SU(2) we have
\begin{equation}
\langle b^j_{mn},b^{j'}_{m'n'}\rangle=\delta^{j,j'}\delta_{mm'}\delta_{nn'}.
\end{equation}
The Hilbert space $L_2(SU(2),d\mu_H)$ decomposes into a direct sum over all inequivalent irreducible representations labelled by $j$
\begin{equation}
\label{jDecompSU2}
L_2(SU(2),d\mu_H)=\bigoplus\limits_j {\cal H}_j\quad\quad{\rm with}\quad {\cal H}_j:=V_j\otimes V^*_j,
\end{equation}
where $V^*_j$ denotes the dual space of $V_j$. A basis in ${\cal H}_j$ is given by $\{b^j_{mn}\, |\, m,n\in -j,-j+1,\cdots, j-1, j\}$.
Now, we will use this fact to construct the spin network basis of ${\cal H}=L_2(\overline{\cal A},d\mu_{AL})$. For this purpose we first consider the Hilbert spaces ${\cal H}_\alpha$ associated with a fixed graph $\alpha$, which can be identified with $L_2({\rm SU(2)}^n,d^n\mu_H)$. For this reason we can construct an orthonormal basis of ${\cal H}_\alpha$ simply by introducing the so called spin network functions
\begin{eqnarray}
\label{SNF}
|s^{\vec{j}}_{\alpha,\vec{n},\vec{m}}\rangle &:&\overline{\cal A}_\alpha\to\mathbb{C},\quad  A\mapsto \langle A\, | \,s^{\vec{j}}_{\alpha,\vec{n},\vec{m}}\rangle \\
\langle A\, |\, s^{\vec{j}}_{\alpha,\vec{n},\vec{m}}\rangle&:=&\sqrt{2j_{e_1}+1}\cdots\sqrt{2j_{e_n}+1}\pi^{j_{e_1}}_{m_{e_1}n_{e_1}}(A(e_1))\cdots
\pi^{j_{e_n}}_{m_{e_n}n_{e_n}}(A(e_n)),\nonumber
\end{eqnarray}
with 
\begin{equation}
\vec{j}:=\{j_{e_1}, \cdots,j_{e_n}\},\quad \vec{m}:=\{m_{e_1},\cdots,m_{e_n}\},\quad \vec{n}:=\{n_{e_1},\cdots,n_{e_n}\}.
\end{equation}
A decomposition in terms of irreducible representations of SU(2) associated with each edge of the graph $\alpha$ is given by
\begin{equation}
{\cal H}_\alpha=\bigoplus\limits_{\vec{j}}{\cal H}_{\alpha,\vec{j}}\quad{\rm with}\quad
{\cal H}_{\alpha,\vec{j}}:=\bigotimes\limits_{i=1}^n{\cal H}_{j_{e_i}}, 
\end{equation}
with ${\cal H}_{j_{e_{i}}}$ defined as in equation (\ref{jDecompSU2}). We choose a fixed set of representations $\vec{j}$ and will discuss how ${\cal H}_{\alpha,\vec{j}}$ can be further decomposed, which will be of advantage when we discuss the solutions to the Gauss constraint later on. Let us choose an arbitrary vertex $v_i\in V(\alpha)$ and consider all edges $\{e_i\}$ intersecting at $v_i$. Let us assume that $\alpha$ has $m=|V(\alpha)|$ vertices. Then we can rewrite ${\cal H}_{\alpha,\vec{j}}$ as
\begin{equation}
{\cal H}_{\alpha,\vec{j}}=\bigotimes\limits_{i=1}^m{\cal H}_{v_i}\quad\quad{\rm with}\quad\quad {\cal H}_{v_i}:=\bigotimes\limits_{\substack{e\in E(\alpha) \\ e_i\cap v_i\not=\emptyset}}{\cal H}_{j_{e_i}}.
\end{equation}
The operators $\hat{Y}^{(v_i,e_i)}_j$ satisfy $[\hat{Y}^{(v_i,e_i)}_j,\hat{Y}^{(v_i,e_i)}_k]=i\epsilon_{jk}^{\quad\ell} \hat{Y}^{(v_i,e_i)}_\ell$, where we have chosen the basis $\{\tau_j\}$ in such a way that $[\tau_j,\tau_k]=\epsilon_{jk}^{\quad\ell}\tau_\ell$. They can be interpreted as components of angular momentum operators. For different edges $e_i\not=e_j$ these operators commute. A natural basis in the context of angular momentum operators is the eigenbasis $\{|jm\rangle\}$, that is labelled by the angular momentum $j$ and the magnetic quantum number $m$. Let us restrict our discussion to the case of one edge first and denote the abstract angular momentum Hilbert space by ${\cal H}^{jm}$ and the associated   spin network Hilbert space for this edge $e$ by ${\cal H}_{jm}$. Then the corresponding spin network functions are
\begin{equation}
\langle A\, |\, j_e m_e\rangle_{n_e}:=\sqrt{2j_e+1}\pi^{j_e}_{m_e n_e}(A(e)).
\end{equation}
For fixed $n$ these states are orthogonal likewise to the angular momentum eigenstates $|j_em_e\rangle$.
Using the definitions of the operators $\hat{Y}^{(v,e)}_j$ in terms of left- and right-invariant vector fields their action on $|jm\rangle_{n}$ is given by
\begin{equation}
\hat{Y}^{(v,e)}_k|j_em_e\rangle_n =\sum\limits_{\tilde{m}_e} \left\{\begin{array}{l} \,\,\,\,i\pi^{j_e}_{m_e\tilde{m}}(\tau_k) \\ -i\pi^{j_e}_{m_e\tilde{m}_e}(\tau_k)  \end{array}\right\}|j_e\tilde{m}_e\rangle_{n_e}.
\end{equation}
In order to rewrite this in terms of standard angular momentum operators $\hat{J}^{(v,e)}_j$ and their eigenbasis $|jm\rangle$ we construct for fixed $n$ a unitary map $W:{\cal H}^{jm}\to {\cal H}_{jm}$ that satisfies $W\hat{J}^{(v,e)}_k W^{-1}=\hat{Y}^{(v,e)}_k$ and is explicitly given by
\begin{equation}
 W:{\cal H}^{jm}\to {\cal H}_{jm}\quad |jm;n\rangle\mapsto W|j_em_e;n_e\rangle=\sum\limits_{\widetilde{m}_e}\pi^j_{m_e\tilde{m}_e}(\epsilon)|j_e\widetilde{m}_e\rangle_{n_e}, 
\end{equation}
with $\epsilon:=i\sigma_2=\begin{pmatrix} 0 & 1 \\ -1 & 0 \end{pmatrix}$.
The inverse map $W^{-1}$ is then just given by
\begin{equation}
 W^{-1}:{\cal H}_{jm}\to {\cal H}^{jm}\quad |j_em_e\rangle_n\mapsto W^{-1}|j_e m_e\rangle_n=\sum\limits_{\widetilde{m}_e}\pi^{j_e}_{m_e\widetilde{m}_e}(\epsilon^{-1})|j_e\widetilde{m}_e; n_e\rangle .
\end{equation}
Now we go back to spin network functions associated with a graph $\alpha$. The discussion above shows that we can apply the unitary map $W$ edgewise and have
\begin{equation}
W^{-1}\pi^{j_e}_{m_en_e}(A(e))=\pi^{j_e}_{m_e\widetilde{m}_e}(\epsilon^{-1})\frac{\langle A\, |\, j_e \tilde{m}_e;n_e\rangle}{\sqrt{2j_e+1}}, 
\end{equation}
here summation over repeated indices is assumed. Hence, the spin network function $|s^{\vec{j}}_{\alpha,\vec{m},\vec{n}}\rangle$ can be rewritten in the abstract angular momentum basis as
\begin{equation}
\langle A |s^{\vec{j}}_{\alpha,\vec{m},\vec{n}}\rangle=\pi^{j_{e_1}}_{m_{e_1}\tilde{m}_{e_1}}(\epsilon^{-1})\langle A \, |\, j_{e_1} \tilde{m}_{e_1};n_{e_1}\rangle\cdots \pi^{j_{e_n}}_{m_{e_n}\tilde{m}_{e_n}}(\epsilon^{-1})\langle A \, |\, j_{e_n} \tilde{m}_{e_n};n_{e_n}\rangle .
\end{equation}
By means of the unitary map $W$ we identify $|s^{\vec{j}}_{\alpha,\vec{m},\vec{n}}\rangle$ with abstract angular momentum states and the operators $\hat{Y}^{(v,e)}_j$ with angular momentum operators $\hat{J}^{(v,e)}_j$ and thus we can also discuss the further decomposition of ${\cal H}_{\alpha,\vec{j}}$ in the context of angular momentum coupling theory. Let introduce the following operator associated with the vertex $v_i$
\begin{equation}
(\hat{J}^{(v_i)})^2:=\eta^{jk}\hat{J}^{(v_i)}_j\hat{J}^{(v_i)}_k\quad\quad{\rm with}\quad \hat{J}^{(v_i)}_j:=\sum\limits_{\substack{e\in E(\alpha) \\ e\cap v_i\not=\emptyset}}\hat{J}^{(v_i,e)}_j, 
\end{equation}
where $\eta^{jk}$ denotes again the Cartan-Killing metric for su(2). For each $v_i$ the operator $(\hat{J}^{(v_i)})^2$ acts only on ${\cal H}_{v_i}$ non trivially and has the eigenvalues $l_{v_i}(l_{v_i}+1)$, where the particular value of $l_{v_i}$ are determined by the values $\{j_{e_i}\}$ associated to the edges, that intersect in $v_i$. $l_{v_i}$ can be interpreted as the total angular momentum to which the individual angular momenta associated to the edges couple to. Hence, given the operators $(\hat{J}^{(v_i)})^2$ at each vertex $v_i$ we can label their associated eigenspaces by $l_{v_i}$ and denote them by ${\cal H}_{\alpha,\vec{j},l_{v_i}}$. Likewise to the decomposition in terms of irreducible representations $\vec{j}$ associated to the edges the Hilbert space ${\cal H}_{\alpha,\vec{j}}$ further decomposes into the following direct sum
\begin{equation}
{\cal H}_{\alpha,\vec{j}}=\bigoplus\limits_{\vec{l}}{\cal H}_{\alpha,\vec{j},\vec{l}}\quad{\rm with}\quad
\vec{l}=(l_{v_1},\cdots,l_{v_m}),\quad {\cal H}_{\alpha,\vec{j},\vec{l}}:=\bigotimes\limits_{i=1}^m {\cal H}_{\alpha,\vec{j},l_{v_i}}.
\end{equation}
Thus, the Hilbert space associated with a given graph $\alpha$ can be rewritten as
\begin{equation}
{\cal H}_\alpha=\bigoplus\limits_{\vec{j},\vec{l}}{\cal H}_{\alpha,\vec{j},\vec{l}}
\end{equation}
and states in this Hilbert space are characterized by the irreducible representations, that are associated to the edges and vertices of the graph. For this reason we can label the spin network functions also by this data yielding $|s_{\alpha,\vec{j},\vec{l}}\rangle$. The difference on the form in (\ref{SNF}) is that here the coupling basis for angular momenta has been used for the Hilbert spaces ${\cal H}_{l_{v_i}}$ whereas in (\ref{SNF}) the product basis was used. In the following sections we will use both notations depending on which one is more suitable in the given situation. Now let us focus our discussion again on the kinematical Hilbert space ${\cal H}=L_2(\overline{\cal A},d\mu_{AL})$. We would like to rewrite ${\cal H}$ as a direct sum of the individual ${\cal H}_\alpha$s. However, here we are faced with the following problem. Given a graph $\alpha$ and a cylindrical function $f_\alpha$ that does not depend on the holonomies of at least one of the edges of $\alpha$. Then this function would also be an element of ${\cal H}_{\widetilde{\alpha}}$ for some $\widetilde{\alpha}$, that has less edges and vertices. Hence, ${\cal H}_\alpha\cap{\cal H}_{\widetilde{\alpha}}\not=0$ and therefore the two spaces are not orthogonal. A similar situation occurs when a function depends on the holonomies of two adjacent edges $e_1,e_2$ in $\alpha$ such that for an edge $\tilde{e}$ in $\widetilde{\alpha}$ we have $\widetilde{e}=e_1\circ e_2$. As a consequence, we have to introduce some further rules on how the irreducible representations are associated to the edges of the graph in order to write ${\cal H}$ as an orthogonal decomposition of the ${\cal H}_\alpha$s. For this purpose we introduce the notion of an admissible labeling of edges and vertices. Given a graph $\alpha$ we call a labeling of the edges and vertices of $\alpha$ by irreducible representations admissible if none of the edges carries a trivial representation and furthermore  no two-valent vertex carries a trivial representation. We denote graph Hilbert spaces with admissible labelings by ${\cal H}'_\alpha$. Then we can rewrite the kinematical Hilbert space for LQG as
\begin{equation}
\label{HkinDecint}
{\cal H}=\bigoplus\limits_\alpha {\cal H}'_\alpha=\bigoplus\limits_\alpha\bigoplus\limits_{\substack{\vec{j},\vec{l} \\ {\rm admissible}}} {\cal H}_{\alpha,\vec{j}\vec{l}}.
\end{equation}
This decomposition will be important in the following section when we discuss the dynamics of loop quantum gravity, that is encoded in the quantum Einstein's equations of loop quantum gravity. 
\section{The Quantum Einstein's Equations of Loop Quantum Gravity}
Following the Dirac quantization program requires in the case of loop quantum gravity to implement the Gauss, diffeomorphism and Hamiltonian constraint as operators on the kinematical Hilbert space ${\cal H}$ introduced in the last section. Let us denote these operators by $\widehat{\cal C}_G(\vec{\Lambda})$, $\vec{\cal C}(\vec{N})$ and $\widehat{{\cal C}}(N)$, the quantum analog of the classical Einstein's equations, the so called quantum Einstein's equation of loop quantum gravity are given by
\begin{equation}
\widehat{\cal C}_G(\vec{\Lambda})\psi_{\rm phys}(A)=0,\quad  \widehat{\vec{\cal C}}(\vec{N})\psi_{\rm phys}(A)=0,\quad \widehat{\cal C}(N)\psi_{\rm phys}(A)=0, 
\end{equation}
where $\psi_{\rm phy}(A)$ denotes the physical states, which live in the physical Hilbert space ${\cal H}_{\rm phys}$. The construction of the latter requires apart from finding the (general) solution to the quantum Einstein's equations also to define an inner product on the set of physical states. In this chapter we will restrict our discussion on the definition and solutions of the Gauss constraints. The remaining diffeomorphism and Hamiltonian constraint will be discussed in detail in the chapter by Laddha and Varadarajan in \cite{Book}.
\subsection{Solutions to the Gauss constraint: Gauge-invariant Spinnetwork Functions}
The Gauss constraint is solved using techniques from ordinary lattice gauge theory, where a similar constraint is involved in the theory. Technically, we have two possibilities to construct the solution space, which we will denoted by ${\cal H}^{\cal G}$. Either we can define an operator $\widehat{\cal C}_G(\vec{\Lambda})$ generating infinitesimal gauge transformation or we can consider the exponentiated version $\hat{U}({\cal C}_G)$, that generates finite gauge transformations. The solution space will be the same in both cases. How the infinitesimal gauge transformations can be implemented in the quantum theory is explained in detail for instance in \cite{Thiemann:2007zz}. Here we will consider finite gauge transformation, implemented by unitary operators. As discussed before the holonomy $A(e)$ transforms under gauge transformation as $A(e)\to A^g(e)=g(b(e))A(e)g^{-1}(e)$. The matrix elements of representations of $A(e)$ have thus the following transformation behavior 
\begin{eqnarray}
\pi^j_{m_en_e}(A(e))\to \pi^j_{m_en_e}(A^g(e))&=&\pi^j_{m_en_e}(g(b(e))A(e)g^{-1}(f(e))\\
 &=&\pi^j_{m_e\alpha_e}(g(b(e)))\pi^j_{\alpha_e\beta_e}(A(e))\pi^j_{\beta_en_e}(g^{-1}(f(e)).\nonumber
\end{eqnarray}
In order to construct gauge invariant spin network functions (SNF), first we write the SNF in (\ref{SNF}) in more compact form as
\begin{eqnarray}
\label{SNFcomp}
|s^{\vec{j}}_{\alpha,\vec{n},\vec{m}}\rangle &:&\overline{\cal A}_\alpha\to\mathbb{C},\quad  A\mapsto \langle A\, | \,s^{\vec{j}}_{\alpha,\vec{n},\vec{m}}\rangle \\
\langle A\, |\, s^{\vec{j}}_{\alpha,\vec{n},\vec{m}}\rangle&:=& \prod\limits_{k=1}^n \sqrt{2j_{e_k}+1}\pi^{j_{e_k}}_{m_{e_k} n_{e_k}}(A(e_k)).\nonumber
\end{eqnarray}
Secondly, for the reason that the gauge transformation act on the beginning and final point only, which are precisely the vertices of the graph,  we rewrite the product of edges occurring above as
\begin{eqnarray}
\label{SNFcompvertex}
\langle A\, |\, s^{\vec{j}}_{\alpha,\vec{n},\vec{m}}\rangle&:=& \prod\limits_{v\in v(\alpha)}\prod\limits_{\substack{e\in E(\alpha)\\ e\cap v\not=\emptyset}} \sqrt{2j_{e}+1}\pi^{j_{e}}_{m_{e} n_{e}}(A(e)).
\end{eqnarray}
Let us consider one individual vertex, at which we have n outgoing edges. For simplicity we will consider only outgoing edges first, but will discuss the more general case below. At the vertex $v$ the SNF transforms under gauge transformation as
\begin{eqnarray}
\langle A^g\, |\, s^{\vec{j}}_{\alpha,\vec{n},\vec{m}}\rangle\Big|_v&=& \prod\limits_{\substack{e\in E(\alpha)\\ e\cap v\not=\emptyset}} \sqrt{2j_{e}+1}
\pi^{j_{e}}_{m_{e} \alpha_{e}}(g(b(e)))
\pi^{j_{e}}_{\alpha_{e} n_{e}}(A(e)).
\end{eqnarray}
Let us denote the tensor product of the Hilbert spaces associated with each edge at $v$ as before by ${\cal H}_{v}=\otimes_{\substack{e\in E(\alpha)\\ e\cap v\not=\emptyset}}{\cal H}_{j_e}$. We can define a basis of ${\cal H}_{v}$ in terms of tensors of type $(0,n)$, denoted by $\{t_i\}$ with components $t_i^{\alpha_1\cdots\alpha_n}$, one index for each representation $j_e$. We can define a dual basis, denoted by $\{\tilde{t}^{i}\}$ with components $\tilde{t}^i_{\alpha_1\cdots\alpha_n}$ associated with ${\cal H}^*_{v}=\otimes_{\substack{e\in E(\alpha)\\ e\cap v\not=\emptyset}}{\cal H}^*_{j_e}$, where each ${\cal H}^*_{j_e}$ carries the dual representation $\overline{\pi}^{j_e}$, by requiring
\begin{equation}
\tilde{t}^j(t_i)=\tilde{t}^j_{\alpha_1\cdots\alpha_n}t_i^{\alpha_1\cdots\alpha_n}=\delta^i_j.
\end{equation}
The gauge transformation act on these tensors and its duals by
\begin{eqnarray}
t_i^{\alpha_1\cdots\alpha_n}\to (t')_i^{\alpha_1\cdots\alpha_n}&=&\pi^{j_{e_1}}(g(v))^{\alpha_{1}}_{\quad\beta_1}\cdots \pi^{j_{e_n}}(g(v))^{\alpha_n}_{\quad\beta_n}t_i^{\beta_1\cdots\beta_n} \\
\tilde{t}^i_{\alpha_1\cdots\alpha_n}\to (\tilde{t}')^i_{\alpha_1\cdots\alpha_n}&=&\overline{\pi}^{j_{e_1}}(g(v))^{\beta_{1}}_{\quad\alpha_1}\cdots \pi^{j_{e_n}}(g(v))^{\beta_n}_{\quad\alpha_n}\tilde{t}^i_{\beta_1\cdots\beta_n} \\
&=&\overline{\pi}^{j_{e_1}}(g^{-1}(v))^{\quad\beta_{1}}_{\alpha_1}\cdots \pi^{j_{e_n}}(g^{-1}(v))^{\quad\beta_n}_{\alpha_n}\tilde{t}^i_{\beta_1\cdots\beta_n},  
\end{eqnarray}
where we have used that the dual representation $\overline{\pi}(g(v))=\pi(g^{-1}(v))^T$ and used the notation $\pi^{j}_{mn}(g(v))=\pi^j(g(v))^m_{\quad n}$. Now we are interested in those tensors which are invariant under gauge transformations, which will be denoted by $\{i^k\}$. In terms of their components gauge invariance means
\begin{equation}
\pi^{j_{e_1}}(g(v))^{\alpha_{1}}_{\quad\beta_1}\cdots \pi^{j_{e_n}}(g(v))^{\alpha_n}_{\quad\beta_n}i_k^{\beta_1\cdots\beta_n}=i_k^{\alpha_1\cdots\alpha_n}
\end{equation}
and likewise for their corresponding dual tensors. An intertwiner $i$ between $m$ dual representations $\overline{\pi}^{j_1},\cdots,\overline{\pi}^{j_m}$ and $n$ representations $\pi^{j_1},\cdots\pi^{j_n}$ is a covariant map
\begin{equation}
i: \bigotimes\limits_{k=1}^m {\cal H}_{j_{e_k}}\to \bigotimes\limits_{\ell=1}^n{\cal H}_{j_{e_\ell}}
\end{equation}
and can also be understood as an invariant tensor in $\bigotimes\limits_{k=1}^m {\cal H}^*_{j_{e_k}}\otimes \bigotimes\limits_{\ell=1}^n{\cal H}_{j_{e_\ell}}$. We will use this fact to construct gauge invariant spin network functions. In our example we have a vertex $v$ with n outgoing edges. We achieve that the spin network is invariant under gauge transformation at $v$ when we contract the SNF with the corresponding intertwiner $i_v$ at $v$, in our example this leads to
\begin{eqnarray}
\left[\langle A\, |\, s^{\vec{j}}_{\alpha,\vec{n},\vec{m}}\rangle\Big|_v \right]_{\rm inv}&=&i_{v\vec{m}}\prod\limits_{\substack{e\in E(\alpha)\\ e\cap v\not=\emptyset}} \sqrt{2j_{e}+1}\pi^{j_{e}}_{\vec{m} n_{e}}(A(e))\\
&=&
 \sqrt{2j_{e_1}+1}\cdots\sqrt{2j_{e_n}+1}
i_v^{m_1\cdots m_n}
\pi^{j_{e_1}}_{m_{1} n_{e_1}}(A(e_1))\cdots\pi^{j_{e_n}}_{m_{n} n_{e_n}}(A(e)).\nonumber
\end{eqnarray}
We generalize our discussion to a vertex, that has $v_m$ ingoing edges and $v_n$ outgoing edges. Again we can construct the gauge invariant part of the SNF at this vertex by contracting with an intertwiner $i_v$, which has components of the form $i^{ m_1\cdots m_{v_n}}_{v \, n_1\cdots n_{v_m}}$. Thus, we can construct an invariant SNF by contracting the gauge variant SNF in (\ref{SNF}) at each vertex with a corresponding intertwiner. We will denote the gauge invariant SNF $|s^{\vec{j}}_{\alpha,\vec{i}}\rangle$, where $\vec{i}\in\{i_v\, |\, v\in V(\alpha)\}$ is the set of intertwiners associated with the graph. The gauge invariant SNF is then given by
\begin{eqnarray}
\label{SNFginv}
|s^{\vec{j}}_{\alpha,\vec{i}}\rangle &:&\overline{\cal A}_\alpha\to\mathbb{C},\quad  A\mapsto \langle A\, | \,s^{\vec{j}}_{\alpha,\vec{i}}\rangle\nonumber \\
\langle A\, |\, s^{\vec{j}}_{\alpha,\vec{i}}\rangle&:=& \prod\limits_{v\in V(\alpha)} i_v \prod\limits_{\substack{e\in E(\alpha)\\ e\cap v\not=\emptyset}} \sqrt{2j_{e}+1}\pi^{j_{e}}(A(e))\nonumber \\
&=&\prod\limits_{v\in V(\alpha)} i^{m_1\cdots m_{v_n}}_{v \, n_1\cdots n_{v_m}} \sqrt{2j_{e_1}+1}\cdots\sqrt{2j_{e_{v_n+v_m}}+1}\\
&& \pi^{j_{e_1}}_{m_1 n_{e_1}}(A(e_1))
\cdots \pi^{j_{e_{v_n}}}_{m_{v_n} n_{e_{v_n}}} (A(e_{v_n}))\pi^{j_{e_{v_n+1}}}_{m_{e_{v_n+1}} n_{1}}(A(e_{v_n+1}))\cdots
\pi^{j_{e_{v_n+v_m}}}_{m_{e_{v_n+v_m}} n_{v_m}}(A(e_{v_n+v_m})).\nonumber
\end{eqnarray}
Here each vertex has $v_n$ outgoing and $v_m$ ingoing edges and we have labeled set of edges $\{e_1,\cdots,c_{v_n+v_m}\}$ in such a way, that $e_1,\cdots, e_{v_n}$ are the outgoing edges and $e_{v_{n+1}},\cdots e_{v_n+v_m}$ are the ingoing edges.
\\
Going back to the decomposition of ${\cal H}$ in (\ref{HkinDecint}), the gauge invariant Hilbert space corresponds to the case where the edges at all vertices couple to a total angular momentum of zero. Thus, we have for the gauge invariant Hilbert space denoted by ${\cal H}^G_{\rm inv}$
\begin{equation}
{\cal H}^G_{\rm inv}=\bigoplus\limits_\alpha\bigoplus\limits_{\substack{\vec{j},\vec{l} \\ {\rm admissible}}} {\cal H}_{\alpha,\vec{j}\vec{l}=0}.
\end{equation}
The Hilbert space ${\cal H}^G_{\rm inv}$ and therefore the solution space of the Gauss constraint is a subspace of the kinematical Hilbert space ${\cal H}$. For the remaining constraints of the quantum Einstein's equations, this will be no longer be the case and the construction of their corresponding solution spaces is more complicated and will be discussed in the chapter by Laddha and Varadarajan in \cite{Book}.
~\\
~\\
In the discussion above we have derived ${\cal H}^G_{\rm inv}$ by starting with the configuration space $\overline{\cal A}$ and implemented the finite gauge transformations on ${\cal H}$. Afterwards the solution space ${\cal H}^G_{\rm inv}$ was constructed as a subspace of ${\cal H}$. Alternatively, one can also obtain ${\cal H}^G_{\rm inv}$ by considering the reduced quantum configuration space $\overline{\cal A}/\overline{\cal G}$, which consists of all generalized connections modulo (generalized) gauge transformations $\overline{\cal G}$. The latter are the extension of the gauge transformations ${\cal G}$ from the classical configuration space ${\cal A}$ to the quantum configurations space $\overline{\cal A}$.  
In this case only gauge invariant cylindrical functions are considered from the beginning and the final Hilbert space one obtains is also ${\cal H}^G_{\rm inv}$. 
\section{Geometric Operators and Their Properties}
\label{Sec:GeomOp}
One of the special properties of the AL-representation used in loop quantum gravity introduced in the last section is that one can define operators corresponding to geometrical objects such as volume, area and length. For the KS--representation it has been shown that geometric operators can be implemented using similar techniques as for the AL-representation \cite{Sahlmann:2010hn}.
This is a consequence of the choice of the particular smearing of the elementary variables discussed above yielding to the holonomy and flux variables. If we had for instance chosen a three dimensional smearing like for the standard Fock quantization, the implementation of these geometrical operators in the quantum theory would not be possible.
\\ 
 Among those geometrical operators the most simple one is the area operator from the point of view of its quantization as well as with regards to the spectrum of these operators, therefore we will discuss this operator first.

\subsection{The Area Operator}
The area operator was first introduced by Smolin \cite{Smolin:1992qz} and then further analyzed by Rovelli and Smolin in the loop representation \cite{VRS}, which is a representation based on loops instead of graphs and that was used in the earlier days of loop quantum gravity. Ashtekar and Lewandowski \cite{Ashtekar:1996eg} discussed the spectrum of the area operator in the connection representation.
In this section we want to discuss the implementation of the area operator as well as its spectrum in detail. At the end of the section we will briefly comment on the volume and length operator.
\\
\\
The strategy one adopts to quantize is the following: As a first step we have to express the classical expression, such as the area, in terms of Ashtekar variables $(A,P)$. Afterwards we need do find a regularization of it, meaning that in our case the  area needs to be written as a function of holonomies and fluxes. The guiding principle for the regularization is, that in the limit where the regulator is removed, the classical area in terms of $(A,P)$ should be recovered. Since corresponding operators for holonomies and fluxes exists, the regularized area can then be promoted to a (regularized) operator on the kinematical Hilbert space ${\cal H}$, whose detailed properties usually still depends on the chosen regularization. In a final step, one has to show that in the limit where the chosen regulator tends to zero  a well defined operator is obtained.
The classical area functional associated to a surface $S$ is given by the following expression
\begin{equation}
A_S=\int\limits_U d^2u\sqrt{\det(X^*q)}(u), 
\end{equation}
where $q$ denotes the ADM 3-metric and $X: U\to S$ is an embedding of the surface. Here $U\subset \mathbb{R}^2$ and $X^*$ denotes the pull back of $X$. 
The coordinates on the embedded surface $S$ are given by the embedding functions $X^a$ with $a=1,2,3$ and let us denote the two coordinates parametrizing the surface by $u_1$ and $u_2$.  Given the embedding we can construct two tangent vector fields on $S$ 
\begin{equation}
X^a_{,u_1}:=\frac{\partial X^a}{\partial u_1},\quad X^a_{,u_2}:=\frac{\partial X^a}{\partial u_2}
\end{equation}
and also a co-normal vector field $n_a$ that is determined from the condition
\begin{equation}
\label{conormal}
n_aX^a_{u_,i}=0\quad{\rm for}\quad i=1,2.
\end{equation}
The determinant in the area functional can be expressed as
\begin{equation}
\det(X^*q)=q_{u_1u_1}q_{u_2u_2} - q_{u_1u_2}q_{u_2u_1}
=\left(X^a_{,u_1}X^b_{,u_1}X^c_{u_2}X^d_{,u_2} - X^a_{,u_1}X^b_{,u_2}X^c_{,u_2}X^d_{,u_1}\right)q_{ab}q_{cd}.
\end{equation}
In order to quantize the area functional we need to express it in terms of Ashtekar variables. For this purpose we consider the expression $\det(q)n_a n_b q^{ab}$ and use that we can express the inverse metric as 
\begin{equation}
q^{ab}=\frac{1}{2}\frac{1}{\det(q)}\epsilon^{acd}\epsilon^{bef}q_{ce}q_{df}.
\end{equation}
Furthermore, we see from (\ref{conormal}) that $n_a=\epsilon_{abc}X^{c}_{,u_1}X^d_{,u_2}$ yielding
\begin{eqnarray}
\label{DetXq}
\det(q)n_an_bq^{ab}
&=&
\det(q)n_an_b\frac{1}{2}\frac{1}{\det(q)}\epsilon^{acd}\epsilon^{bef}q_{ce}q_{df}\nonumber \\
&=&
\epsilon_{ak\ell}X^k_{,u_1}X^\ell_{,u_2}\epsilon_{bmn}X^m_{,u_1}X^n_{,u_2}\frac{1}{2}\epsilon^{acd}\epsilon^{bef}q_{ce}q_{df}\nonumber \\
&=&
q_{u_1u_1}q_{u_2u_2} - q_{u_1u_2}q_{u_2u_1}.
\end{eqnarray}
The inverse metric has a simple form in Ashtekar variables given by $q^{ab}=\frac{1}{k\gamma}P^a_jP^b_k \delta^{jk}/ \det(P)$ and depends only on the densitized triad. From $P^a_j=k\gamma\sqrt{\det(q)}e^a_j$ we get $\det(q)=k^3\gamma^3\det(P)$  yielding
\begin{equation}
\det(q)q^{ab}=k^2\gamma^2P^a_jP^b_k \delta^{jk}
\end{equation}
from which we can conclude using (\ref{DetXq}) that
\begin{equation}
\sqrt{\det(X^*q)}=k\gamma\sqrt{n_an_bP^a_jP^b_k \delta^{jk}}=k\gamma\sqrt{P^{\perp}_jP^{\perp}_k \delta^{jk}}, 
\end{equation}
where $P^\perp_j$ denotes the projection of $P^a_j$ in normal direction with respect to the surface. Note that often one chooses the basis $\tau_j:=-i\sigma_j/2$ in su(2) with $\sigma_j$ being the Pauli matrices for which the Cartan-Killing metric on su(2) $\eta_{jk}$ becomes 
$\eta_{jk}:=Tr(ad(\tau_j)ad(\tau_k))=-2\delta_{jk}$ and then one uses the Killing metric in the expression above and adjusts the pre-factors accordingly. 
\\
 In order to quantize the area functional we need to choose a regularization of the classical expression. For this purpose, we choose a family of non-negative densities $f^\epsilon_u(u')$ on the surface $S$ as regulators, which tend to $\delta_u(u')$ in the limit $\epsilon\to 0$, that is
\begin{equation}
\lim_{\epsilon\to 0} f^\epsilon_u(u')=\delta_x(y), 
\end{equation}
where $\delta_u(u')$ is the delta-function on $S$ peaked at $u$. Given $f^\epsilon_u(u')$ we can define a regularized version of $P^\perp_j(u)$ denoted by $[P^\perp_j]^\epsilon$ and defined as
\begin{equation}
\label{RegP}
[P^\perp_j]^\epsilon(u):=\int\limits_S d^2u' f^\epsilon_u(u')P^\perp_j(u').
\end{equation}
In the limit where the regulator is removed we have
\begin{equation}
\lim_{\epsilon\to 0} [P^\perp_j]^\epsilon(u)=P^\perp_j(u).
\end{equation}
Using $[P^\perp_j]^\epsilon$ and a point-splitting, a common technique used in quantum field theory, we can define a regularized expression for the area functional as
\begin{eqnarray}
[A_S]^\epsilon&:=&k\gamma \int\limits_S d^2u \left(\int\limits_S d^2 u'\int\limits_S d^2u'' f^\epsilon_u(u')P^\perp_j(u') f^\epsilon_u(u'')P^\perp_k(u''')\delta^{jk}\right)^{\frac{1}{2}}\nonumber\\
&=&k\gamma \int\limits_S d^3u \sqrt{[P^\perp_j]^\epsilon(u)[P^\perp_k]^\epsilon(u)\delta^{jk}}.
\end{eqnarray}
Obviously, we have $\lim_{\epsilon\to 0}[A_S]^\epsilon=A_S$ in the classical theory. To define a regularized area operator $[\hat{A}_S]^\epsilon$ we use the following strategy: We replace $P^\perp_j(u')$ in (\ref{RegP}) by the operator $\hat{P}^\perp_j:=-i\hbar\frac{\delta}{\delta A^j_\perp}$ yielding a regularized operator $[\hat{P}^\perp_j]^\epsilon$. Afterwards we have to compute the action of $[\hat{P}^\perp_j]^\epsilon$ on spin network functions and check whether $[\hat{P}^\perp_j]^\epsilon$ yields a well defined operator. This is indeed the case and one obtains
\begin{equation}
[\hat{P}^\perp_j]^\epsilon(u) |s^{\vec{j}}_{\alpha,\vec{m},\vec{n}}\rangle=\frac{\hbar}{2}\sum\limits_{v\in V(\alpha)}f^\epsilon_u(v)\sum\limits_{\substack{e\in E(\alpha)\\ e\cap v\not=\emptyset}} \kappa(S,e)\hat{Y}_j^{(v,e)}|s^{\vec{j}}_{\alpha,\vec{m},\vec{n}}\rangle, 
\end{equation}
where the operators $\hat{Y}_j^{(v,e)}$ have been defined in (\ref{Yoperator}). Hence, we can rewrite the regularized area operator in the following form
\begin{equation}
[A_S]^\epsilon |s^{\vec{j}}_{\alpha,\vec{m},\vec{n}}\rangle=4\pi\gamma\ell_p^2 \int\limits_S d^2u \sqrt{\left(\sum\limits_{v\in V(\alpha)}f^\epsilon_u(v)\sum\limits_{\substack{e\in E(\alpha)\\ e\cap v\not=\emptyset}} \kappa(S,e)\hat{Y}_j^{(v,e)}\right)^2}|s^{\vec{j}}_{\alpha,\vec{m},\vec{n}}\rangle, 
\end{equation}
where we have used the definition of the Planck length $\ell_p=\hbar G_N=\frac{8\pi}\hbar k$. Next we choose $\epsilon$ sufficiently small such that for a given $u\in S$ $f^\epsilon_u(v)$ is non-vanishing only for at most one vertex $v$. Thus, we have $f^\epsilon_u(v)f^\epsilon_u(v')=\delta_{v,v'}(f^\epsilon_u(v))^2$ and we obtain
\begin{equation}
[A_S]^\epsilon |s^{\vec{j}}_{\alpha,\vec{m},\vec{n}}\rangle=4\pi\gamma\ell_p^2 \int\limits_S d^2u
\sum\limits_{v\in V(\alpha)}f^\epsilon_u(v)
\sqrt{\left(\sum\limits_{\substack{e\in E(\alpha)\\ e\cap v\not=\emptyset}} \kappa(S,e)\hat{Y}_j^{(v,e)}\right)^2}|s^{\vec{j}}_{\alpha,\vec{m},\vec{n}}\rangle .
\end{equation}
As a final step we have to remove the regulator yielding to a well defined area operator $\hat{A}_S$ on the kinematical Hilbert space ${\cal H}$ of the form
\begin{eqnarray}
[A_S] |s^{\vec{j}}_{\alpha,\vec{m},\vec{n}}\rangle &:=&
\lim_{\epsilon\to 0}[A_S]^\epsilon |s^{\vec{j}}_{\alpha,\vec{m},\vec{n}}\rangle \\
&=&4\pi\gamma\ell_p^2 \int\limits_S d^2u \sum\limits_{v\in V(\alpha)} \delta_u(v)
\sqrt{\left(\sum\limits_{\substack{e\in E(\alpha)\\ e\cap v\not=\emptyset}} \kappa(S,e)\hat{Y}_j^{(v,e)}\right)^2}|s^{\vec{j}}_{\alpha,\vec{m},\vec{n}}\rangle .\nonumber
\end{eqnarray}
From the above expression for the area operator we realize that in the sum over all vertices of the graph $\alpha$ only those vertices will contribute which are intersection points of the surface $S$ as otherwise $\kappa(S,e)=0$. For this reason we can write the area operator in more compact form by introducing the set 
$I(S)$ of intersection points of edges of type up and type down, that is given by
\begin{equation}
I(S)=\{v\in e\cap S | \kappa(S,e)\not=0, \,\, e\in E(\alpha)\}.
\end{equation}
This yields to the final form of the area operator that we will use in the following
\begin{equation}
\label{areaOPYx}
\hat{A}_S|s^{\vec{j}}_{\alpha,\vec{m},\vec{n}}\rangle=4\pi\gamma\ell_p^2\sum\limits_{v\in I(S)}\sqrt{\big(\sum\limits_{e\,\text {at}\, v}\kappa(S,e)\hat{Y}^{(e,v)}\big)^2}|s^{\vec{j}}_{\alpha,\vec{m},\vec{n}}\rangle .
\end{equation}
 Let us now discuss the spectrum of the area operator. At each intersection point $v\in I(S)$ we have edges of type up, edges of type down and edges of type in that will not contribute to the spectrum. In order to write the expression under the square root in (\ref{areaOPYx}) in compact form we introduce the following operators:
\begin{equation}
\hat{Y}_j^{v,u}:=\sum\limits_{e\in E(v,u)} \hat{Y}_j^{(v,e)}\quad \hat{Y}_j^{v,d}:=\sum\limits_{e\in E(v,d)} \hat{Y}_j^{(v,e)}.
\end{equation}
Here $E(v,u), E(v,d)$ denotes all edges of type up and down respectively that intersect each other in the point $v$. Then we have for each intersection point $v$
\begin{eqnarray}
\label{Yud}
\left(\sum\limits_{\substack{e\in E(\gamma)\\ e\cap v\not=\emptyset}}\kappa(S,e) \hat{Y}_j^{(v,e)}\right)^2
&=&
\left(\hat{Y}^{v,u} - \hat{Y}^{v,d}\right)^2 \nonumber \\
&=&(\hat{Y}^{v,u})^2 +  (\hat{Y}^{v,d})^2 - 2\hat{Y}^{v,u}\hat{Y}^{v,d} \nonumber \\
&=&2(\hat{Y}^{v,u})^2 +  2(\hat{Y}^{v,d})^2 - (\hat{Y}^{v,u}+\hat{Y}^{v,d})^2.
\end{eqnarray}
We used in the second line that $[\hat{Y}_j^{v,u},\hat{Y}_k^{v,d}]=0$. Furthermore, the operators $(\hat{Y}^{v,u})^2$, $(\hat{Y}^{v,d})^2$ and $ (\hat{Y}^{v,u}+\hat{Y}^{v,d})^2$ mutually commute. Moreover, we choose an explicit basis $\tau_j=-i\sigma_j/2$ for which the operators $\hat{Y}^{(v,e)}$ satisfy the usual angular momentum algebra given by
$[\hat{Y}^{(v,e)}_i,\hat{Y}^{(v,e)}_j]=\epsilon_{ijk}\hat{Y}^{(v,e)}_k$. Then we have that the operators $(\hat{Y}^{(v,e)})^2\equiv \delta^{jk}\hat{Y}^{(e,v)}_j\hat{Y}^{(v,e)}_k$ locally act as
\begin{equation}
-\delta^{ij} R_i R_j=-\langle{R},{R}\rangle\equiv -\Delta_{SU(2)}, \text{ or }-\delta^{ij} L_i L_j=-\langle{L},{L}\rangle\equiv -\Delta_{SU(2)},
\end{equation}
where $-\Delta_{SU(2)}$ is the positive definite SU(2) Laplacian with spectrum $j(j+1)$, due to our choice of basis for su(2). Hence, the same holds for the operators $(\hat{Y}^{v,u})^2$, $(\hat{Y}^{v,d})^2$, and $(\hat{Y}^{v,u}+\hat{Y}^{v,d})^2$, they act as Laplacians in the respective direct sum of representations. Therefore the spectrum of the operators involved in (\ref{Yud}) can be easily computed and we obtain 
 \begin{equation}
{\rm Spec}(\hat{A}_S)=4\pi\gamma\ell_p^2\sum\limits_{v\in I(S)}
\sqrt{2j_{u,v}(j_{u,v}+1) +  2j_{d,v}(j_{d,v}+1) -j_{u+d,v}(j_{u+d,v}+1)}.
\end{equation}
Here $j_{u,v},j_{d,v}$ denote the total angular momentum of the edges of type up (down respectively) at the intersection point $v$ and $j_{u+d,v}$ total coupled angular momentum of the up and down edges whose values range between $|j_{u,v}-j_{d,v}|\leq j_{u+d,v}\leq j_{u,v}+j_{d,v}$.
Let us consider the eigenvalue at one intersection point $v$. The smallest possible eigenvalue that we can get  occurs when either $j_{u,v}=0$ and $j_{d,v}=\frac{1}{2}$ or vice versa. The eigenvalue denoted by $\lambda_0$ is non vanishing and given by
\begin{equation}
\lambda_0=2\pi\gamma\ell_p^2\sqrt{3}
\end{equation}
and is known as the area gap in loop quantum gravity. The area gap plays an important role in the description of black hole physics within loop quantum gravity and  black hole entropy calculations can be used the fix the value of the Immirzi parameter $\gamma$ as discussed in the chapter by Barbero and Perez in \cite{Book}.
\\
\subsection{The Volume Operator}
The volume operator enters crucially into the construction of the dynamics of the quantum Einstein's equations for the reason that the classical co-triad is expressed as the Poisson bracket between the connection and the  classical volume functional using the Thiemann identity (see the chapter by Laddha and Varadarajan in \cite{Book}).
In the case of the area operator, the area functional depends on the momenta $P^a_j$ only, which is also true for the classical volume functional, that for a given region $R$ in the spatial manifold $\Sigma$ reads
\begin{equation}
V_R=\int\limits_R d^3x \sqrt{\det(q)}=(k\gamma)^{\frac{3}{2}}\int\limits_R d^3x \sqrt{|\det(P^a_j)|}.
\end{equation}
Likewise to the case of the area operator we need to choose a regularization of $V_R$ in order to write the volume functional in terms of fluxes $P(S,f)$ for which well defined operators exist yielding as a first step a regularized expression of the classical volume functional $V_R$. For the latter it is natural to choose a partition ${\cal P}^\epsilon$  of the spatial region $R$ in terms of cubic cells $C^\epsilon$ and adapted 2-surfaces for each cubic cell. For this purpose we introduce a coordinate system $(x^a)$ and assume that each $C^\epsilon$ has a volume of less than $\epsilon$ in the chosen coordinate system and that two different cells share only points on their boundaries. For each cubic cell $C^\epsilon$ we introduce three 2-surfaces $\{S_a\,|\, a=1,2,3\}$ chosen in a way such that the coordinates components $x^a$ are constant along $S_a$ for $a=1,2,3$ following the notation in \cite{Ashtekar:2004eh}. Furthermore, each $S_a$ has the property that it divides $C^\epsilon$ into two disjoint parts. We can now use the surfaces $\{S_a\,|\, a=1,2,3\}$ to formulate a regularized volume functional denoted by $V^\epsilon_R$ as a function of fluxes over the surfaces $\{S_a\,|\, a=1,2,3\}$. Going back to the definition of the flux $P(S,f)$ in (\ref{DefP(S,f)}) we will choose as the smearing functions $f^j$ the su(2) basis elements $\tau_j$ and define $P_j(S):=P(S,\tau_j)$. Given this, we have
\begin{equation}
V_R^\epsilon=(k\gamma)^\frac{3}{2}\sum\limits_{C^\epsilon \in P^{\epsilon}}\sqrt{|Q_{C^\epsilon}|}
\end{equation}
with
\begin{equation}
Q_{C^\epsilon}:=\frac{1}{3!}\epsilon^{jk\ell}\epsilon^{abc}P_j(S_a)P_k(S_b)P_\ell(S_c).
\end{equation}
In the classical theory we have $\lim_{\epsilon\to 0} V^\epsilon_R=V_R$, however in the quantum theory the removal of the regular has to be taken with more care. While in the case of the area operator after the regulator has been removed the final operator does not depend on the chose background structure of the regularization a different situation occurs for the volume operator. In the case of the volume operator once the regulator is removed, the resulting operator still depends on the chosen partition and thus carries a memory of the chosen regularization. As a consequence this operator depends on the chosen background structure we have chosen during the regularization procedure and therefore the limit does not yield an appropriate candidate for a volume operator because it fails to be covariant under spatial diffeomorphisms. This problem can be circumvented by first averaging over the possible background structures, whose dependence enters into the volume operator in a rather simple way, before removing the regulator. The requirements that we obtain a well defined operator when the regulator is removed as well as that the final operator is covariant under spatial diffeomorphisms, are restrictive enough to uniquely determine the final form of the operator up to a global constant, that we will denote regularization constant $c_{\rm reg}$ in the following. 
 ~\\
~\\
 In the literature two different volume operators exist, one introduced by Rovelli and Smolin (RS) \cite{VRS} and one introduced by Ashtekar and Lewandowski (AL) \cite{VAL}, which come out of a priori equally justified but different regularization techniques. In the classical theory both regularized versions, the RS- as well as the AL-volume -although being of different kind- yield the classical volume functional once the regulator is removed. However, in the quantum theory the removal of the regulator is more subtle and this is the reason why one ends up with two different quantum operators.
 Both volume operators act non-trivially only on vertices where at least three edges intersect. At a given vertex the operators have the following form
\begin{eqnarray}
\label{eq:VolOp}
\hat{V}_{v,\rm RS}&=&c_{\rm RS}\sum\limits_{e_I\cap e_J\cap e_K=v}\sqrt{\big|\hat{Q}_{IJK}\big|}\nonumber \\
\hat{V}_{v,\rm AL}&=&c_{\rm AL}\sqrt{\Big|\sum\limits_{e_I\cap e_J\cap e_K=v}\epsilon(e_I,e_J,e_K)\hat{Q}_{IJK}\big|}.
\end{eqnarray}
Here $\hat{Q}_{IJK}:=\epsilon^{ijk}\hat{Y}_i^{(v,{e_I})}\hat{Y}_j^{(v,{e_J})}\hat{Y}_k^{(v,{e_K})}$ is an operator involving only flux operators and thus right and left  invariant vector fields and $c_{\rm RS},c_{\rm AL}$ are regularization constants. The sum runs over all ordered triples of edges intersecting at the vertex $v$.  A detailed discussion about the regularization of the volume operator can for instance be found in \cite{Ashtekar:2004eh,Thiemann:2007zz}.
The main differences between these two operators is that the RS-operator is not sensitive to the orientation of the triples of edges and is therefore covariant under homeomorphisms. The AL-operator has likewise to the $\kappa(S,e)$ in the area operator a similar sign factor $\epsilon(e_I,e_J,e_K)$ that can take the values $\{+1,0,-1\}$ and is the sign of the cross product of the tangent vectors at $v$ of the triple of edges $e_i,e_j,e_k$ that intersect at this vertex $v$. Furthermore, the sum over triples of edges involved in both operators occurs outside the square root in case of the RS and inside the square root in case of the AL-operator. Due to the sign factor $\epsilon(e_I,e_J,e_K)$ the operator $\hat{V}_{\rm AL}$ is covariant only under diffeomorphisms.
\\ 
The spectral analysis of the volume operator is more complicated than for the area operator and can in general not be computed analytically.  A general formula for the computation of matrix elements of the AL-volume operator has been derived in \cite{JBVol}. Those techniques have been used to analyze the spectrum of the volume operator numerically up to a vertex valence of 7 in a series of papers \cite{JBDR}. Their work showed that the spectral properties of the volume operator depend on the embedding of the vertex that enters via the sign factors $\epsilon(e_i,e_j,e_k)$ into the construction of the AL-operator. Particularly, the presence of a volume gap, that is a smallest non-vanishing  eigenvalue, depends on the geometry of the vertex. A consistency check for both volume operators has been discussed in \cite{CCVol} where the Thiemann trick, discussed in detail in  the chapter by Laddha and Varadarajan in \cite{Book}, has been used to define an alternative flux operator. The alternative flux operator is then compared to the usual flux operator and consistency of both operators could for instance fix the undetermined regularization constant $c_{\rm AL}=\ell_p^3/\sqrt{48}$ in the volume operator. Furthermore, the RS-operator did not  pass this consistency check and the reason that it worked for the AL-operator is exactly the presence of those sign factors $\epsilon(e_I,e_J,e_K)$ in the AL-operator. 
\\
\\
A technique to compute matrix elements of the volume operator with respect to semiclassical states analytically has been developed in \cite{Giesel:2006um}. This method relies on the idea of an expansion of the matrix elements of the volume operator in a power series of matrix elements of operators, that can be computed analytically. These operators in the expansion are chosen in such a way that the error caused by this expansion can be estimated and can be well controlled.
\subsection{The Length Operator}
A length operator for LQG was introduced in \cite{Tlength}. The length operator is in some sense the most complicated one among the kinematical geometrical operators. Let us recall the the length of a curve $c: [0,1]\to\Sigma$ classically is given by
\begin{equation}
\label{cllen}
\ell(c)=\int\limits_{0}^1\sqrt{q_{ab}(c(t))\dot{c}^a(t)\dot{c}^b(t)}dt=\int\limits_{0}^1\sqrt{e^i_{a}(c(t))e^j_b(c(t))\dot{c}^a(t)\dot{c}^b(t)\delta_{ij}}dt, 
\end{equation}
here $\dot{c}^a$ denotes the components of the tangent vector associated to the curve.
When we express the metric $q_{ab}$ in terms of Ashtekar variables we obtain
\begin{equation}
q_{ab}=\frac{k}{4}\epsilon_{acd}\epsilon_{bef}\epsilon_{ijk}\epsilon^{imn}\frac{P^c_jP^d_kP^m_eP^n_f}{\det(P)}, 
\end{equation}
which is a non-polynomial function in terms of the electric fields and therefore a regularization in terms of flux operators similar to the area and volume operator does not exist.  Furthermore, the denominator being the square of the volume density cannot be defined on a dense set in ${\cal H}$ because it has a huge kernel.
One possibility to quantize the length used in \cite{Tlength} is to use for the co-triads that occur in (\ref{cllen}) the Thiemann trick and replace them by a Poisson bracket between the connection and the volume functional.  
This yields a length operator that involves a square root of two commutators between holonomy  operators along the curve $c$ and the volume operator. In this way the inverse volume density can be avoided and the volume occurs only linearly in the commutator. Also, the length operator does not change the graph or the spin labels of the edges likewise to the area and volume operator.
However, since the length operator becomes even a function of the volume operator its spectral analysis becomes even more complicated than for the volume operator itself and very little about the spectrum of the length operator is known except for low valence vertices.
\\
Another length operator was introduced in \cite{Blength} where the Thiemann trick was not used for the quantization. The regularization adapted in \cite{Blength} is motivated from the dual picture of quantum geometry and uses that the curve can be expressed as an intersection of two surfaces. This allows to express the tangent vector of the curve in terms of the normals of the surfaces. The inverse volume issue discussed above is circumvented by using a Tikhonov regularization for the inverse RS-volume-operator. For this length operator the spectral properties have only be analyzed for a vertex of valence 4, which is monochromatic, that is all spins are identical.
Anonther alternative length operator for LQG has been discussed in \cite{Mlength} where a different regularization has been chosen such that the final length operator can be expressed in terms of other geometrical objects the area, volume and flux operators. In this work the AL-operator is used and the inverse volume operator is also defined using a Tikhonov regularization similar to the one in \cite{Blength}.
\section{Summary}
In this review we presented a brief introduction to the kinematical setup of loop quantum gravity. Loop quantum gravity can be understood as a framework for canonically quantizing general relativity. This approach leads to a quantum theory based on quantum geometry for the reason that not only the matter part of the theory but also the geometry itself is quantized. In section 1 we briefly mentioned earlier attempts to canonically quantize general relativity using ADM-variables. However, these approaches could only provide a quantum theory that was constructed at a rather formal level since neither the functional analytical details about the kinematical Hilbert space had been worked out nor could the dynamics of the quantum theory be implemented rigorously. 
But precisely the quantization of the constraints that encode the dynamics of the quantum theory needs to be understood in great detail if one wants to analyze characteristic properties and consequences of quantum geometry. Particularly, the Hamiltonian constraint is a non-polynomial function of the elementary phase space variables in contrast to the Hamiltonian used in other gauge theories in the context of the standard model of particle physics.
Nevertheless, these earlier results were important because they already showed what kind of complications occur if one tries to carry over the standard quantization used in ordinary quantum mechanics to general relativity. Progress regarding this aspect was made when the connection variables were introduced by Ashtekar\cite{Ashtekar:1986yd} leading to a reformulation of general relativity in terms of an SU(2) gauge theory as discussed in section 2. As a consequence it involves next to the the spatial diffeomorphism and the Hamiltonian constraint also known from the ADM-formalism an additional SU(2) Gauss constraint.
Although, the Hamiltonian constraint keeps its non-polynomial form also with respect to these new variables the advantage of the connection formulation is, that general relativity can be formulated in the language of ordinary gauge theories. This leads to a form of the constraints in the new variables that looks much closer to what we are familiar with from other gauge theories. Therefore, techniques developed in those fields could be taken as a point of reference for constructing the quantum theory underlying loop quantum gravity. Taking this into account the choice of holonomies and fluxes as presented in section 3 is a very natural choice as elementary phase space variables for the theory. We introduced the notion of cylindrical functions and flux vector fields acting on them in order to give a precise definition of the holonomy--flux algebra used in loop quantum gravity. The choice of the classical algebra and its related properties are important in the sense that the corresponding quantum theory will of course depend on the particular choice because we obtain the quantum theory by finding representations of the underlying classical algebra. In the case of the holonomy--flux algebra the first representation that was found is the Ashtekar--Lewandowski \cite{Ashtekar:1991kc, Ashtekar:1994mh} representation discussed in section 4. Interestingly, later the LOST-theorem \cite{Lewandowski:2005jk}  proved that this is the only representation of the holonomy--flux algebra if the symmetries of the theory, particularly the spatial diffeomorphisms are taken very seriously. Other representation that violate one of the assumptions used in the LOST-theorem were found by Sahlmann and Koslowski \cite{Koslowski:2011vn}. We finished section 4 by introducing spin networks which provide an orthornormal basis for the kinematical Hilbert space. Beside being a very useful tool as far as computations in loop quantum gravity are concerned they also deliver  insight into the question how quantum states look like in loop quantum gravity. Each spin network is defined on a graph that consists of a finite number of edges that are one--dimensional objects embedded into the spatial manifold we obtained from 3+1 split. These edges are labeled with so called spin quantum numbers and the vertices of the graph carry intertwiners. These data can be understood as describing a particular state of quantum geometry at the kinematical level and by varying these data we would obtain different states of quantum geometry. To go beyond the kinematical level we have to consider the dynamics of quantum geometry that is described by the quantum Einstein's equations. These are the classical analogue of Einstein's equations in general relativity. In the context of Dirac quantization for constrained systems the formulation of the quantum Einstein's equations requires to implement the classical constraints as operators on the kinematical Hilbert space. If one considers a reduced phase space quantization approach for loop quantum gravity\cite{AQGIV}, then formulating the dynamics requires to define a (physical) Hamiltonian on the physical Hilbert space. The latter  is obtained by quantizing directly the reduced phase space. A more detailed presentation of the quantum dynamics can be found in the chapter by Laddha and Varadarajan in \cite{Book}. In section 5 we only start to introduce the topic of quantum dynamics and we restrict our discussion to the construction of solutions to the Gauss constraint only. The corresponding solutions are gauge invariant spin network functions and the remaining dynamical operators associated with finite spatial diffeomorphisms and the infinitesimal Hamiltonian constraint are well defined on the gauge invariant Hilbert space. We finished this article with a brief review on geometrical operators. These are operators associated with geometrical quantities like length, area and volume. That these operators can be implemented is a special property of the kinematical representation used in loop quantum gravity and related to the fact that holonomies as well as fluxes are used as the elementary variables. In a Fock representation, used in ordinary quantum field theory, those operators are not well defined. At the kinematical level the spectrum of the area operator can be computed analytically and interestingly it turns out to be discrete and a smallest non-vanishing eigenvalues exist a so called area gap. For the volume and length operator the complete spectrum is still unknown but one has analyzed the volume operator for special spin networks states with low valence \cite{JBDR}.
\\
\\
The kinematical setup introduced in this review provides the mathematical foundation for most of the research done in loop quantum gravity. In the context of loop quantum cosmology, that is a symmetry reduced model for loop quantum gravity and introduced in the chapter by Agullo and Singh in \cite{Book}, the kinematical representation discussed here is adopted and specialized to the context of cosmological models. Also the particular implementation of the quantum Einstein's equations discussed in the chapter by Laddha and Varadarajan in \cite{Book} is closely related to the choice of the kinematical representation. In the context of black hole physics the area operator plays an important role and provides new insights on a quantum mechanical description of the black hole entropy as discussed in the chapter by Barbero and Perez in \cite{Book}. Furthermore, a motivation for spin foam models, which aim to provide the corresponding covariant formulation of loop quantum gravity in the context of path integral quantization, is again the kinematical framework presented in this chapter. Therefore, also in the covariant approach the kinematical Hilbert space plays an important role. More details on the covariant approach can be found in the chapters by Bianchi, Dittrich and Oriti in \cite{Book}.
 
\end{document}